\documentclass[a4paper]{jpconf}

\usepackage{graphicx}
\usepackage{dcolumn}
\usepackage{bm}
\usepackage[square,sort,comma,numbers]{natbib}
\usepackage{graphicx}
\usepackage{mathtools}
\usepackage{float}
\usepackage{textcomp}
\usepackage{xcolor}
\usepackage{cancel}
\begin{document}
\title{Measurement of the electric quadrupole amplitude in atomic thallium $6P_{1/2}\rightarrow6P_{3/2}$ transition using electromagnetically induced transparency}

\author{Wei~Ling~Chen, Wei~Min~Hsu, Li~Bang~Wang and Yi~Wei~Liu$^*$}

\address{$^1$ Department of Physics, National Tsing Hua University, Hsinchu 30013, Taiwan}
\address{$^2$ Center for Quantum Technology, National Tsing Hua University, Hsinchu 30013, Taiwan}

\ead{ywliu@phys.nthu.edu.tw}

\begin{abstract}
We report a measurement of the transition amplitude ratio $\chi$ of an electric quadrupole ($E2$) to a magnetic dipole ($M1$) of the $6P_{1/2}\rightarrow6P_{3/2}$ transition in atomic thallium. 
We utilized the electromagnetically induced transparency (EIT) mechanism and the sideband-bridging technique to resolve the isotopic transitions and their hyperfine manifold. 
Our measurement gave $\chi_{205}~=~0.2550(20)(7)$ for $^{205}$Tl, and $\chi_{203}=0.2532(73)(7)$ for $^{203}$Tl, which was measured for the first time. 
Our result provides a reference point for the theoretical calculation of atomic structure and new input for the long dispute on the atomic thallium PNC measurements.
\end{abstract}

\section{Introduction}
The atomic parity non-conservation (PNC) experiment makes a significant contribution to testing the standard model (SM) in a low energy regime. 
By exchanging Z bosons between bound electrons and nuclear quarks in the electroweak interaction, the PNC effect increases rapidly with nuclear charge, so the PNC measurement in heavy atoms has become one of the most promising approaches to search for physics beyond SM \cite{bouchiat1974weak,bouchiat2011atomic,sandars1987parity}.

The atomic PNC signal originates from the nuclear spin-independent (NSI) and the nuclear spin-dependent (NSD) parts. The NSI part, which is the dominant contribution to PNC \cite{dzuba2012revisiting,porsev2009precision,geetha1998nuclear}, depends on the nuclear weak charge \cite{ginges2004violations,sahoo2021new}.
The NSD part of the PNC provides the value of the anapole moment of a nucleus \cite{flambaum1997anapole,flambaum1984nuclear,dzuba2011calculation,dmitriev2004p}.
This value puts a constraint on the PNC meson-nucleon coupling constant \cite{haxton2002nuclear,flambaum1997anapole}. 

All the atomic PNC experiments require high-precision calculations in atomic structure.
The most accurate measurement to date has been achieved with Cs by the Wieman's group \cite{wood1997measurement} in 1997.
Alkali atoms have the advantage of being with a better understood wave function for the highly accurate calculation.
This experiment has reached an unprecedented 0.35$\%$ accuracy, which could be compared with the prediction of the SM \cite{dzuba2002high}.

In addition, several heavier ions such as $\rm{Yb}^+$, $\rm{Ba}^+$, and $\rm{Ra}^+$, having a chain of isotopes, are also considered potential candidates for exploring the anapole moment through the PNC effect and measurements \cite{dzuba2011calculation,sahoo2011parity,sahoo2011parity2}.

Another important atomic PNC measurement has been achieved with atomic thallium in 1995 by the Oxford group \cite{edwards1995precise} and the Seattle group \cite{vetter1995precise}, using the optical rotation spectroscopy.
Due to the more complicated atomic structure in comparison with cesium, the thallium PNC reached 1\% \cite{vetter1995precise} experimental accuracy, but the theoretical calculation was only 3\% \cite{dzuba1987calculation}.
Although the uncertainty of thallium PNC is larger, it provides alternative constraints on both the NSI and NSD parts of PNC \cite{haxton2001atomic} and plays a unique role in today's testing of SM in low energy. 
However, the latest atomic thallium PNC measurements were performed more than two decades ago, and they were in great disagreement with each other.

The electric dipole amplitude induced by PNC ($E1_{PNC}$) is one of the PNC observables and can be detected via interference of the $E1_{PNC}$ amplitude and the magnetic dipole ($M1$) amplitude among the same fine structure states, e.g. optical rotation.
In the previous atomic thallium PNC data analysis \cite{majumder1999measurement}, the ratio of the $E1_{PNC}$ to the $M1$ amplitude, $R_{PNC}=~\Im[E1_{PNC}/M1]$, had a $2.2\sigma$ discrepancy between the Oxford group \cite{edwards1995precise} with $R_{PNC}=-15.71(45)\times 10^{-8}$ and the Seattle group \cite{vetter1995precise} with $R_{PNC}=-14.71(17)\times 10^{-8}$. 
Although these two groups used different techniques to calibrate their measured PNC rotations, this discrepancy was still partially attributed to that their optical rotation profile analysis used a different $\chi$ value, which is the ratio of electric quadrupole ($E2$) amplitude to the $M1$ amplitude and defined by the operators $Q^{(2)}$ and $\mu^{(1)}$ \cite{majumder1999measurement}:
\begin{equation}
    \chi\equiv\frac{\omega}{4\sqrt{3}}\frac{\langle J'\vert\vert Q^{(2)}\vert\vert J\rangle}{\langle J'\vert\vert\mu^{(1)}\vert\vert J\rangle}.
\end{equation}

In the last decades, atomic structure and its $\chi$ value have been further researched \cite{safronova2005excitation, porsev2012electric, tang2018relativistic, maartensson1995magnetic, garstang1964transition, warner1968transition, neuffer1977calculation, biemont1996forbidden} with the transition probabilities of forbidden lines.
In 1964, \cite{garstang1964transition} used intermediate coupling theory to obtain the atomic structure.
Following the astrophysical interest, the semi-empirical methods and Scaled-Thomas-Fermi wave functions were used in \cite{warner1968transition}. 
Additionally, \cite{neuffer1977calculation} also calculated the transition probabilities, where the valence-electron wave functions are generated as numerical solutions of the Dirac equation in a modified Tietz central potential.
Then, \cite{maartensson1995magnetic} utilized a relativistic coupled-cluster approach to obtain the atomic wave functions in 1995.
One year later, the energy levels and radiative transition probabilities were calculated by \cite{biemont1996forbidden} with the relativistic Hartree-Fock method, including configuration interaction terms.
For the latest calculation in 2005, the energies of states in Tl are obtained using the third-order relativistic many-body perturbation theory (MBPT) and the SD all-order method \cite{safronova2005excitation}, in which single and double (SD) excitations of the Dirac-Fock (DF) wave functions are summed to all orders.
The transition rates for electric-quadrupole and magnetic-dipole transitions in Tl were also calculated in the SD approximation.
Besides, the recent references \cite{porsev2012electric, tang2018relativistic}, which studied those approximations in the above methods, also supported the calculation of the transition probabilities.
For the Oxford group, $\chi=0.254$ was used from \cite{neuffer1977calculation}; for the Seattle group, $\chi=0.240$ was derived from their experiment, and this value was also verified subsequently by \cite{safronova2005excitation,maartensson1995magnetic}.
This was one of the important causes leading to the discrepancy in $R_{PNC}$ between these two groups.

The atomic thallium PNC optical rotation experiment is to detect the rotation of laser polarization \cite{wolfenden1991observation,macpherson1991precise,meekhof1993high} propagating through a thallium vapor \cite{vetter1995precise}. 
In such an experiment, $\chi$ plays an essential role in understanding the optical rotation lineshape to derive a reliable value of $R_{PNC}$.
The total optical rotation $\phi$ is contributed by:
\begin{equation}
    \phi(\nu)=\phi_P(\nu)+\phi_F(\nu)+\phi_B(\nu)
\end{equation}
where $\nu$ is the optical frequency, $\phi_P$ is the PNC optical rotation, $\phi_F$ is the Faraday rotation, and $\phi_B$ is the background rotation.
The PNC optical rotation interferes with all the other effects.
The PNC amplitude, $R_{PNC}$, is related to the PNC optical rotation by \cite{vetter1995precise}:
\begin{equation}
\label{phip}
    \phi_P(\nu)=-4\pi L\nu c^{-1}[n(\nu)-1]R_{PNC}
\end{equation}
where $n(\nu)$ is the refractive index and $L$ is the path length.

The $\chi$ value, which is related to the state-mixing Faraday rotation, has great importance in calculating the Faraday rotation.
To acquire an accurate PNC optical rotation, the background rotation and the Faraday rotation need to be subtracted.
For the background rotation, using a dummy tube could remove the background \cite{wolfenden1991observation}; 
For the Faraday rotation, a zero magnetic field could eliminate the first-order Faraday rotation which is proportional to the absorption depth and the external magnetic field \cite{majumder1999measurement,edwards1995precise}. 
However, a small component of the high order Faraday rotation is still generated by the hyperfine-state mixing \cite{cronin1999new}. 
For a transition between states $|i\rangle$ and $|j\rangle$, this high order Faraday rotation is written in the following terms \cite{edwards1995precise}:
\begin{equation}
\label{Faraday}
    \phi_F(\nu)\propto\sum_{F_iF_j}[K_1(F_i,F_j)+\chi^2K_2(F_i,F_j)]L_G(\nu_{ji}-\nu)
\end{equation}
where the angular coefficients $K_1$ and $K_2$ are calculated from the Clebsch–Gordan coefficients and can be found in \cite{edwards1995precise}, and the $L_G$ is a line-shape function. 

\begin{figure}[htbp]
\centering
\includegraphics[width=300 pt]{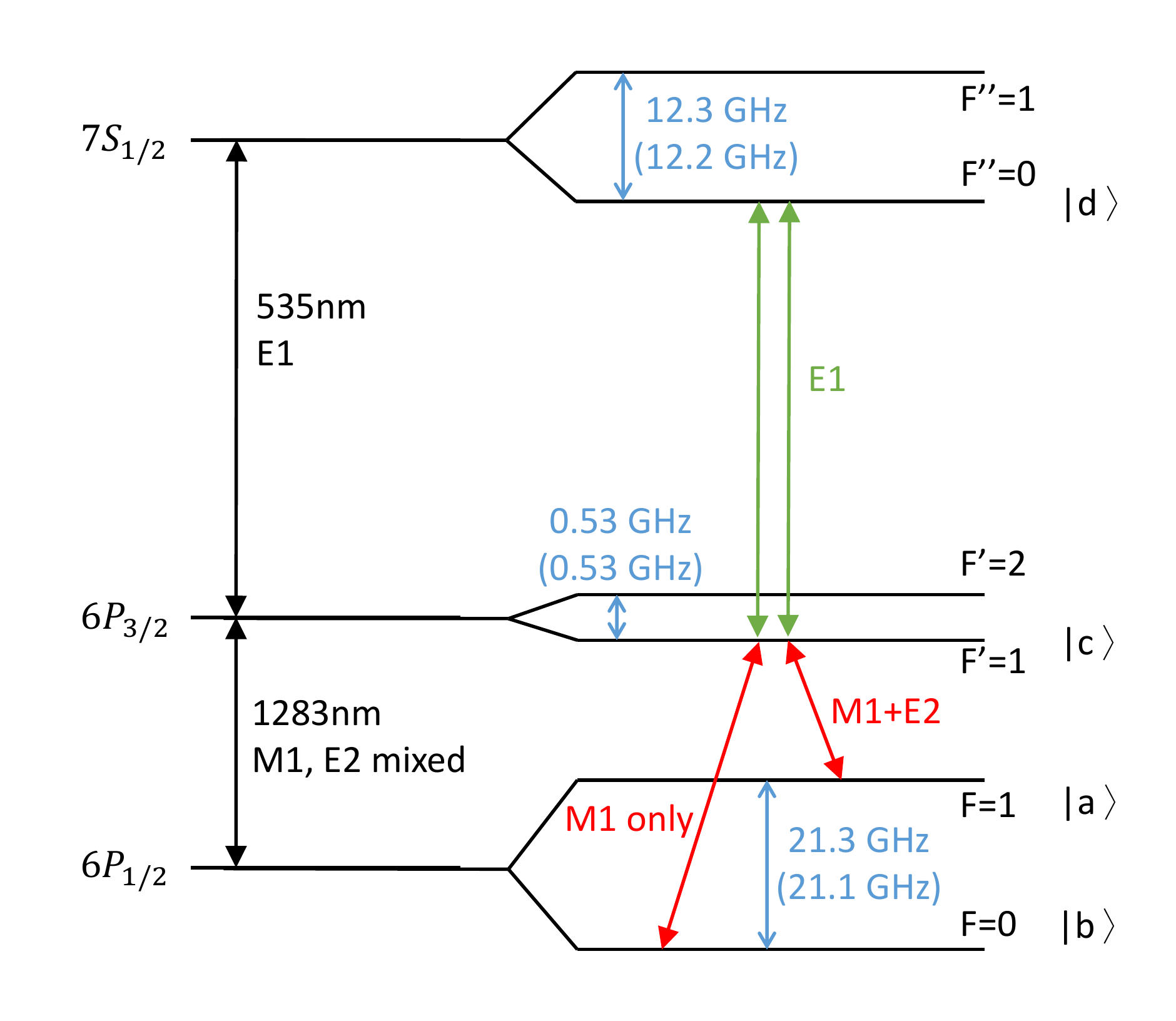}
\caption{The partial energy level diagram of atomic thallium. $F=1\rightarrow F'=1\rightarrow F''=0$ (channel~A) and $F=0\rightarrow F'=1\rightarrow F''=0$ (channel~B) are two channels to measure the $\chi$. The $F=1\rightarrow F'=1$ is mixed with $M1$ and $E2$, and the $F=0\rightarrow F'=1$ is purely $M1$.
The hyperfine splittings of each state for $^{205}$Tl ($^{203}$Tl) are listed in the figure.}
\label{structure}
\end{figure}

In the previous experiment \cite{majumder1999measurement}, the high order electromagnetic transitions ($M1$ \& $E2$) of the $6P_{1/2}\rightarrow6P_{3/2}$ transition were measured by the absorption spectroscopy of $F=1\rightarrow1,2$ and $F=0\rightarrow1,2$ to derive the ratio, $\chi=E2/M1$, as Fig.~\ref{structure} shows in the partial energy level diagram of atomic thallium. 
The $\chi$ value in this early experiment was measured to be 0.2387(10)(38).
The systematic uncertainty was caused by the lineshape uncertainty, which was due to the Doppler broadening and the nearby $^{203}$Tl isotopic transitions, because the individual transition profiles could not be fully resolved. 
Therefore, a method utilizing the Electromagnetically-Induced-Transparency (EIT) to reduce the transition linewidth and prevent one transition from being disturbed by other transitions was proposed by \cite{cronin1998studies}. 

In addition, the EIT method makes the isotopic measurement of the $\chi$ value possible. 
As discussed in \cite{dzuba1986enhancement,pollock1992atomic}, the factor of the atomic structure would be largely canceled in the ratio measurement of the relative PNC between isotopes.
The isotopic PNC ratio is independent of the theoretical calculation. 
In comparison with a single isotope, the measurements in a chain of isotopes can reach higher precision. 
Some recent issues \cite{safronova2018search,antypas2019isotopic} have also exhibited the potential in Yb, which has a chain of stable isotopes.
Therefore, it is not constrained by the knowledge of the atomic wave function and has a great potential for searching for new physics beyond the SM \cite{fortson1990nuclear,brown2009calculations}.
 
The ladder-type EIT in atomic thallium could also improve the PNC measurement as \cite{cronin1998studies} demonstrated and discussed. 
Furthermore, it is the technique to resolve the isotope shift and can be applied to the isotopic PNC measurement ($^{205}$Tl or $^{203}$Tl) by selecting the corresponding coupling transition. 
Our experiment presented not only a new $\chi$ value for thallium $6P_{1/2}\rightarrow6P_{3/2}$ transition but also a development for future EIT-PNC measurements.

\section{EIT for $\chi$}

EIT is an optical technique referring to the suppression upon a resonant absorption using a second laser field (coupling beam). 
It can be utilized for sub-Doppler spectroscopy to measure $\chi$. 
Due to the sub-Doppler linewidth of EIT, it can resolve the overlapping Doppler-broadened isotopic transitions.
Additionally, by modulating the coupling beam, the differential output of the demodulating lock-in amplifier largely suppresses the background noise by the common-mode rejection.

This method significantly reduces the interference between two isotopes, $^{203}$Tl (29.5\%) and $^{205}$Tl (70.5\%).
In the previous experiment \cite{majumder1999measurement}, the direct Doppler-broadened absorption spectroscopy was employed, and two different transitions were overlapped together because the linewidth was as large as $\sim400$~MHz.
The EIT technique narrowed the profile linewidth down to 75~MHz, which is much smaller than the $6p_{3/2}$ state hyperfine splitting, $531(30)$~MHz \cite{shie2013frequency}. 

In our experiment, we used two EIT channels to perform the signal amplitude ratio measurement, from which the $\chi$ value could be derived.
These two channels are composed of a common coupling transition and two different forbidden transitions.
Furthermore, a high frequency fiber-coupled electro-optic modulator (f-EOM) was utilized to bridge the large frequency gap between the $6P_{1/2}$ hyperfine splitting, $21.30957(35)$~GHz \cite{chen2012absolute}, in order to reduce the possible systematic errors caused by a large frequency scanning.

\subsection{Typical three-level EIT model}

Considering a typical three-level atomic system, including $\vert1\rangle$ (ground state), $\vert2\rangle$ (middle state), and $\vert3\rangle$ (excited state) in a ladder-type configuration, the density-matrix equations of motion under the rotating-wave approximation can be written as \cite{joshi2003electromagnetically}:
\begin{equation}
\begin{split}
\label{density matrix equation}
\dot{\rho_{\rm{11}}}=&~2\gamma_{\rm{2}}\rho_{\rm{22}}+\imath\Omega_{\rm{1}}(\rho_{\rm{12}}-\rho_{\rm{21}}),\\
\dot{\rho_{\rm{22}}}=&-2\gamma_{\rm{2}}\rho_{\rm{22}}+2\gamma_{\rm{3}}\rho_{\rm{33}}-\imath\Omega_{\rm{1}}(\rho_{\rm{12}}-\rho_{\rm{21}})+\imath\Omega_{\rm{2}}(\rho_{\rm{23}}-\rho_{\rm{32}}),\\
\dot{\rho_{\rm{12}}}=&-(\gamma_{\rm{2}}-\imath\Delta_{\rm{1}})\rho_{\rm{12}}+\imath\Omega_{\rm{2}}\rho_{\rm{13}}+\imath\Omega_{\rm{1}}(\rho_{\rm{11}}-\rho_{\rm{22}}),\\
\dot{\rho_{\rm{13}}}=&-(\gamma_{\rm{3}}-\imath(\Delta_{\rm{1}}+\Delta_{\rm{2}}))\rho_{\rm{13}}+\imath\Omega_{\rm{2}}\rho_{\rm{12}}-\imath\Omega_{\rm{1}}\rho_{\rm{23}},\\
\dot{\rho_{\rm{23}}}=&-(\gamma_{\rm{2}}+\gamma_{\rm{3}}-\imath\Delta_{\rm{2}})\rho_{\rm{23}}-\imath\Omega_{\rm{1}}\rho_{\rm{13}}+\imath\Omega_{\rm{2}}(\rho_{\rm{22}}-\rho_{\rm{33}}).\\
\end{split}
\end{equation}
$2\Omega_1$, derived from the E1 dipole approximation, and $2\Omega_2$, derived from the higher order (M1+E2) transition moment, are the Rabi frequencies for the probe ($\vert1\rangle\rightarrow\vert2\rangle$) and the coupling ($\vert2\rangle\rightarrow\vert3\rangle$) transitions, respectively.
$\Delta_1$ is the detuning of the probe and $\Delta_2$ is the detuning of the coupling. $\gamma_i$~'s are the decay rates for each state. 
Without any buffer gas, the collision time between thallium atoms is 0.49 ms, which is much longer than the 7.5~ns natural lifetime of the $7S_{1/2}$ state.
The collision will also quench the population of the $6P_{3/2}$ metastable state. 
Thus, the broadenings of both the coupling and probe transition are equivalent to the lifetime effects.
We then take the population relaxation rate ($2\gamma_i$) to be two times of the dephasing rate ($\gamma_i$). 
In our experiment, the relevant hyperfine states of atomic thallium are substantially separated, fig. 1, in comparison with the laser linewidths, therefore, all the off-resonant excitations are negligible.

In the steady state, the time derivative is taken to be zero for all the equations above. 
The weak probe approximation assumes that $\Omega_{\rm{1}}$ compared to $\Omega_{\rm{2}}$ is weak in this model, and $\rho_{\rm{11}}\sim1$. 
The steady solution of the off-diagonal term, $\rho_{\rm{21}}$, is \cite{khan2016role,zhu1996sub}:
\begin{equation}
\label{rho12}
\rho_{\rm{21}}=-\frac{\imath\Omega_{\rm{1}}\rho_{\rm{11}}}{\gamma_{\rm{2}}-\imath\Delta_{\rm{1}}+\frac{{\Omega_{\rm{2}}}^2}{\gamma_{\rm{3}}-\imath(\Delta_{\rm{1}}+\Delta_{\rm{2}})}}.
\smallskip
\end{equation}
The complex susceptibility, $\eta(\omega_p)$, for the probe beam can be obtained from the polarizability and is related to the off-diagonal term of the density matrix in the probe transition, $\rho_{\rm{21}}$ \cite{gea1995electromagnetically}:
\begin{equation}
    \begin{split}
        P=~&\frac{1}{2}\epsilon_0E_p[\eta(\omega_p)e^{-\imath\omega_pt}+\rm{c.c.}]\\
        =~&-2\hbar\Omega_1N{E_p^{-1}}\rho_{21}e^{-\imath\omega_pt}+\rm{c.c.}\\
    \end{split}
\end{equation}
where $N$ is the number density of atoms.
Considering the Doppler broadening in the configuration of the probe-coupling counter-propagating, the Doppler-broadened complex susceptibility by integrating all over the entire velocity distribution is written as:
\begin{equation}
\label{eta}
    \eta(\Delta_1,\Delta_2)=\int_{-\infty}^{\infty}\frac{4\imath\hbar{\Omega_1}^2}{\epsilon_0{E_p}^2}\left[\gamma_{\rm{2}}-\imath\left(\Delta_{\rm{1}}+\frac{\omega_p\nu}{c}\right)+\frac{{\Omega_{\rm{2}}}^2}{\gamma_{\rm{3}}-\imath(\Delta_{\rm{1}}+\Delta_{\rm{2}})-\imath(\omega_p-\omega_c)\nu/c}\right]^{-1}N(\nu)d\nu
\end{equation}
$\eta(\Delta_1,\Delta_2)$ is proportional to ${\Omega_1}^2$, and its imaginary part represents the absorption.
Thus, the absorption signal, $\Sigma_{ab}$, is written as:
\begin{equation}
\begin{split}
\label{s}
    \Sigma_{ab}=I_p\cdot\operatorname{Im}[\eta(\Delta_1,\Delta_2)]&=I_p\bm{d}^2\cdot L(\Delta_1,\Delta_2;\Omega_2)\\
    &=S\cdot L(\Delta_1,\Delta_2;\Omega_2)\\
\end{split}
\end{equation}
where $I_p~({E_p}^2)$ is the probe beam intensity, $\bm{d}~=2{\Omega_1}/{E_p}$ is the transition moment of $\vert1\rangle\rightarrow\vert2\rangle$, $L(\Delta_1,\Delta_2;\Omega_2)$ is the EIT profile, and $S$ is the signal amplitude that is proportional to $I_p \bm{d}^2$.
The integration of Eq.~(\ref{eta}) has been analytically calculated by \cite{gea1995electromagnetically} and was used as our EIT profile fitting model with $\Omega_2=$ constant and $\Delta_2=0$.
In this paper, our interest is in measuring $\bm{d}$.

\subsection{EIT in thallium forbidden transition}

For an $E1$-forbidden transition, considering the high order moments $M1$ and $E2$, the transition moment including all the Zeeman sublevels, such as the $6P_{1/2}\rightarrow6P_{3/2}$ of thallium, can be written as \cite{cronin1999new}:
\begin{equation}
\label{s2}
    \bm{d}^2(J,F,m,J',F',m',\theta)=\sum_{m,m'}{\vert p+\chi q\vert}^2\vert\langle J'\vert\vert\mu^{(1)}\vert\vert J\rangle\vert^2.
\end{equation}
Both $p(J,F,m,J',F',m',\theta)$ and $q(J,F,m,J',F',m',\theta)$ are related to the quantum numbers of the states involved in the probe transition. 
$\theta$ is the relative polarization angle between the coupling and the probe beams. 
It affects the relative transition rates through the different Zeeman sublevels of the intermediate state, $\vert2\rangle$.
$\chi$ is the ratio of the $E2$ to $M1$ transition moment as we mentioned earlier. 

In this paper, we focus on two three-level ladder-type EIT transitions of the atomic thallium, belonging to the hyperfine manifold of $6P_{1/2}\rightarrow6P_{3/2}\rightarrow7S_{1/2}$ as Fig.~\ref{structure} shows. 
The two channels are $\vert\rm{a}\rangle\rightarrow\vert\rm{c}\rangle\rightarrow\vert\rm{d}\rangle (channel~A)$ and $\vert\rm{b}\rangle\rightarrow\vert\rm{c}\rangle\rightarrow\vert\rm{d}\rangle (channel~B)$. 
In channel~A, the probe transition ($\vert\rm{a}\rangle\rightarrow\vert\rm{c}\rangle$, $\bm{d}_{\rm{A}}$) is a mixture of $M1$ and $E2$ because $\Delta F=0$. 
In channel~B, the probe transition ($\vert\rm{b}\rangle\rightarrow\vert\rm{c}\rangle$, $\bm{d}_{\rm{B}}$) is $M1$ only because $\Delta F=1$, and therefore $q=0$.
Using Eq.~(\ref{s2}), the complete forms of $\bm{d}_{\rm{A}}$ and $\bm{d}_{\rm{B}}$ are \cite{cronin1999new,cronin1998studies}:

\begin{equation}
\label{cossin}
\begin{split}
  {\bm{d}_{\rm{A}}}^2&=\vert\langle J'\vert\vert\mu^{(1)}\vert\vert J\rangle\vert^2\cos^2\theta(1-3\chi/\sqrt{5})^2/4\\
  {\bm{d}_{\rm{B}}}^2&=\vert\langle J'\vert\vert\mu^{(1)}\vert\vert J\rangle\vert^2\sin^2\theta\\
\end{split}
\end{equation}
where $\langle J'\vert\vert\mu^{(1)}\vert\vert J\rangle$ is the $M1$ transition moment of $6P_{1/2}\rightarrow6P_{3/2}$.
The $6P_{3/2}\rightarrow7S_{1/2}$ coupling transition ($F'=1\rightarrow F''=0$) is $E1$-allowed and shared by both channels. 
Because the same coupling transition, the ${\Omega_{\rm{c}}}^2(={\Omega_2}^2)$, $\gamma_{\rm{d}}(=\gamma_3)$, and $\Delta_{\rm{c}}(=\Delta_2)$ are the same for both channels, and the lineshape part, $L(\Delta_1,\Delta_{\rm{c}};\Omega_{\rm{c}})$, in Eq.~(\ref{s}) can be eliminated in a differential measurement.
From Eq.~(\ref{cossin}), the ratio of the absorption signals between channel~A ($S_{\rm{A}}$) and channel~B ($S_{\rm{B}}$) can be written as:
\begin{equation}
\label{ratio}
    \frac{S_{\rm{A}}}{S_{\rm{B}}}=\frac{{\bm{d}_{\rm{A}}}^2}{{\bm{d}_{\rm{B}}}^2}=(1/4)|(3\chi/\sqrt{5})-1|^2\cot^2\theta.
\end{equation}
By measuring the ratio of the amplitudes between channel~A and B under various relative coupling-probe polarization angles $\theta$, the $\chi$ can be extracted.

Our experiment was based on the differential measurement between channel~A and channel~B. 
While the measurements on these two channels were performed, maintaining the experimental conditions to be the same would be crucial for a high common mode rejection rate.
Thus, we have taken advantage of the high frequency f-EOM to generate two identical sidebands and to bridge the large frequency gap between the two channels.

\section{Experiment}

\subsection{Fiber-EOM frequency bridging}

The hyperfine structure of $6P_{1/2}$ is $\sim21.3$~GHz, which is too large to cover in one continuous scan for our 1283~nm external cavity diode laser (ECDL). 
Furthermore, including power, pointing, and polarization, the accompanied variances became serious systematic errors in our differential measurement.
Therefore, we used a f-EOM to generate two laser sidebands on the probe beam to measure the two EIT signals in a minimum time gap.

\begin{figure}
\centering
\includegraphics[width=320 pt]{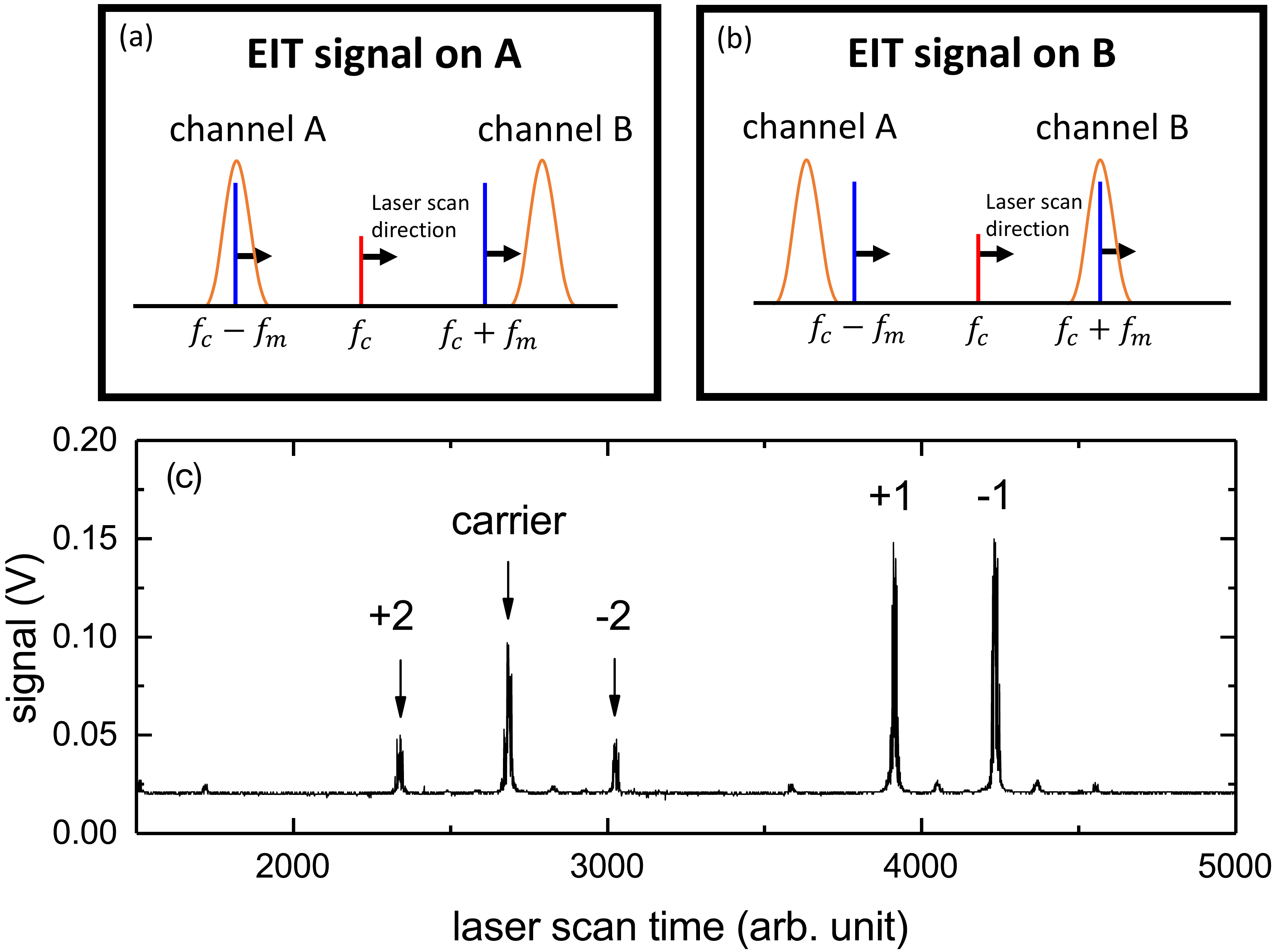}
\caption{The relationship between the f-EOM modulated 1283~nm laser frequencies and two EIT transitions. 
(a) One of the laser sideband positions at the channel~A resonance. The orange lines are the EIT transmission windows with a 21.3~GHz frequency gap. 
The red line is the laser carrier, and the blue lines are the first order laser sidebands. 
The coupling beam is on resonance. While the low frequency sideband of the probe beam scans across the transition $6P_{1/2}(F=1)\rightarrow6P_{3/2}(F'=1)$, the channel~A signal is observed.
(b) The other laser sideband position at the channel~B resonance. While the high frequency sideband of the probe beam scans across the transition $6P_{1/2}(F=0)\rightarrow6P_{3/2}(F'=1)$, the channel~B signal is observed.
(c) The f-EOM modulated the 1283~nm laser to generate the carrier, the first-order sidebands, and the second-order sidebands which were monitored by a reference cavity with a FSR = 1.4~GHz while the 1283~nm laser carrier was scanning.}
\label{carriersidebands}
\end{figure}

The frequency difference between the two sidebands (20.74~GHz) was slightly lower than the hyperfine splitting of the ground state $6P_{1/2}$ to avoid overlapping. 
Fig.~\ref{carriersidebands}~(a) and (b) illustrate the relative positions of two 1283~nm laser sidebands while the sidebands scanned across channel~A and channel~B, respectively. 
The channel~A signal is observed when the low frequency sideband crosses the transition $6P_{1/2}(F=1)\rightarrow6P_{3/2}(F'=1)$;
The channel~B signal is observed when the high frequency sideband crosses the transition $6P_{1/2}(F=0)\rightarrow6P_{3/2}(F'=1)$. 
The AC stark shift \cite{grimm2000optical} of the probe transition, induced by the nearby carrier component and the other sidebands, was estimated to be 78~Hz for these forbidden transitions and is negligible in comparison with the observed linewidth.
With the help of two sidebands, we can successively measure the channel~A and channel~B signals within a few seconds.
That is, the laser frequency itself (carrier) was only scanned 1.1~GHz instead of 21~GHz.
This advantage ensured the minimized variations of the probe and the coupling during the period of acquiring these two channels.
Fig.~\ref{carriersidebands}~(c) shows the carrier and sidebands observed using a reference cavity with 1.4~GHz free spectral range (FSR). 
The amplitude of $\pm1$ sidebands was guaranteed to be identical by the f-EOM. 
The residual amplitude modulation (RAM) has been minimized by careful alignment of the polarization to the input polarization maintained input fiber axis, although this high frequency (21~GHz) RAM has no effect on our measurement.
The difference in the refractive indices (dispersion) between the two sidebands ($\Delta f=21$~GHz) is only $7.02\times10^{-7}$.
The second-order sidebands were also observed.
However, they did not affect our experiment because of being far away from any transitions.

\subsection{Experimental setup}

\begin{figure*}[ht]
\centering
\includegraphics[width=450 pt]{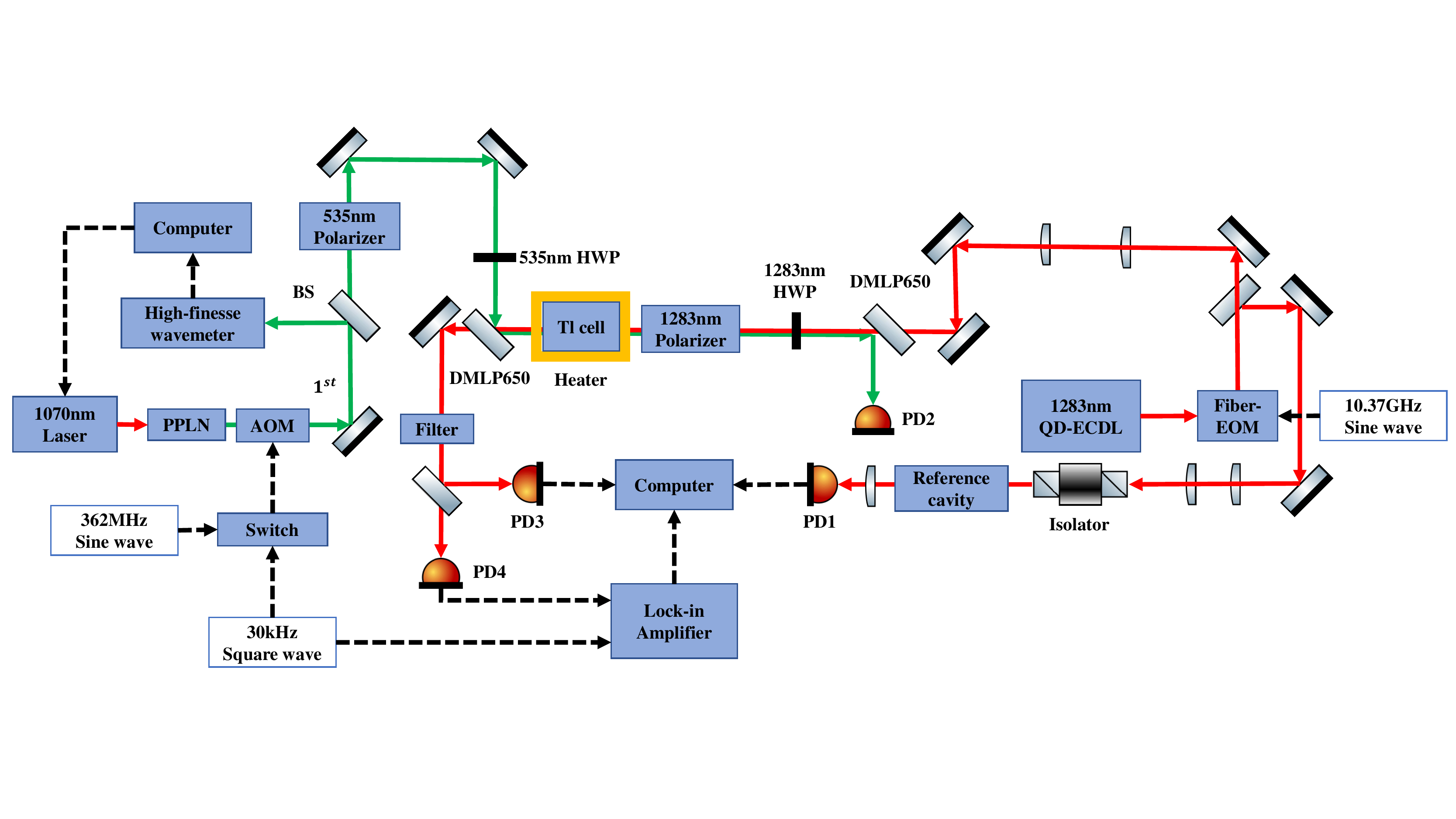}
\caption{Experimental setup. AOM: acousto-optic modulator. PPLN: periodically poled lithium niobate. BS: beam splitter. HWP: half wave plate. QD: quantum dot. ECDL: external cavity diode laser. DMLP650: 650~nm longpass dichroic mirror. 
PD1: photodetector for reference cavity. PD2: photodetector for detecting the coupling beam power. PD3: photodetector for detecting the probe beam power. PD4: photodetector for measuring the EIT signals. 
The green line is the 535~nm laser light. The red line is the  1283~nm laser light. The black dash line is the electronic signal connection.}
\label{expsetup}
\end{figure*}

The schematic diagram of our experimental setup is shown in Fig.~\ref{expsetup}, which is composed of three parts: the coupling laser, the probe laser, and the EIT spectrometer.

The coupling beam (535~nm) was from an amplified 1070~nm ECDL (Newport, TLB-6300-LN) and frequency doubled by a MgO doped periodically poled lithium niobate (HC Photonics, PPLN). 
The temperature of the PPLN was controlled at $103.67~^\circ$C to reach the quasi-phase matching condition. 
An output of 50~mW at 535~nm could be generated from a 1.3~W 1070~nm fundamental laser. 
The acousto-optic modulator (AOM) with a 362~MHz injection frequency was ON/OFF modulated for switching the coupling laser at 30~kHz.
Part of the light after the AOM was directed to the high finesse wavemeter (HighFinesse, WSU-30) to measure the frequency in 1~MHz resolution, and this signal was used for computer feedback lock to stabilize the frequency of the coupling laser.

The probe laser was a 1283~nm quantum dot ECDL \cite{chen2017sub}. It was capable of scanning over 7~GHz without mode hopping. 
The 5~mW output was modulated using the f-EOM, which was injected with a 10.37~GHz microwave.
The microwave was generated using three frequency doubling stages from a frequency synthesizer (HP, 8648B) with an output of 1.29625~GHz. 
The reference cavity with FSR = 1.4~GHz was used for diagnosis.
The optical isolator was essential for the QD-ECDL, which was very sensitive to optical feedback.

The EIT spectrometer was arranged as a coupling-probe counter-propagation configuration with precise polarization control.
The two polarizers for the coupling beam (Thorlabs, GT5-A) and the probe beam (Thorlabs, GT5-C) are Glan-Taylor calcite polarizers with $10^5$ extinction ratio. 
The half-wave plate (HWP) for the coupling beam was used to adjust the coupling beam polarization angle relative to the probe beam. 
The HWP for the probe beam was used to optimize the polarization for better efficiency of the dichroic mirror (Thorlabs, DMLP650), which was used to split or combine the coupling and the probe beams.

Both beams passed through a 10~cm thallium cell at 577~$^\circ$C.
In the interaction region, the probe beam diameter is $\sim$~1.5~mm, and the coupling beam diameter is $\sim$~3~mm.
Due to the modulation transfer mechanism, the 30 kHz modulation was transferred to the unmodulated probe beam. 
The lock-in amplifier (SRS, SR830) was to demodulate the probe beam signal with 30~mV sensitivity and 30~ms time constant.
Therefore, the EIT signals, which were the result of the two-photon interaction between the probe and the coupling beams, could be observed.

\subsection{Single photon direct absorption spectroscopy}

The single-photon direct absorption for both the probe transition $(6P_{1/2}\rightarrow6P_{3/2})$ and coupling transition $(6P_{3/2}\rightarrow7S_{1/2})$ was performed as the preparation stages before acquiring the EIT spectrum. 
In these experiments, we tested our laser systems and characterized several important features of our apparatus.

\subsubsection{1283~nm $E1$-forbidden probe transition}

\begin{figure}
\centering
\includegraphics[width=260 pt]{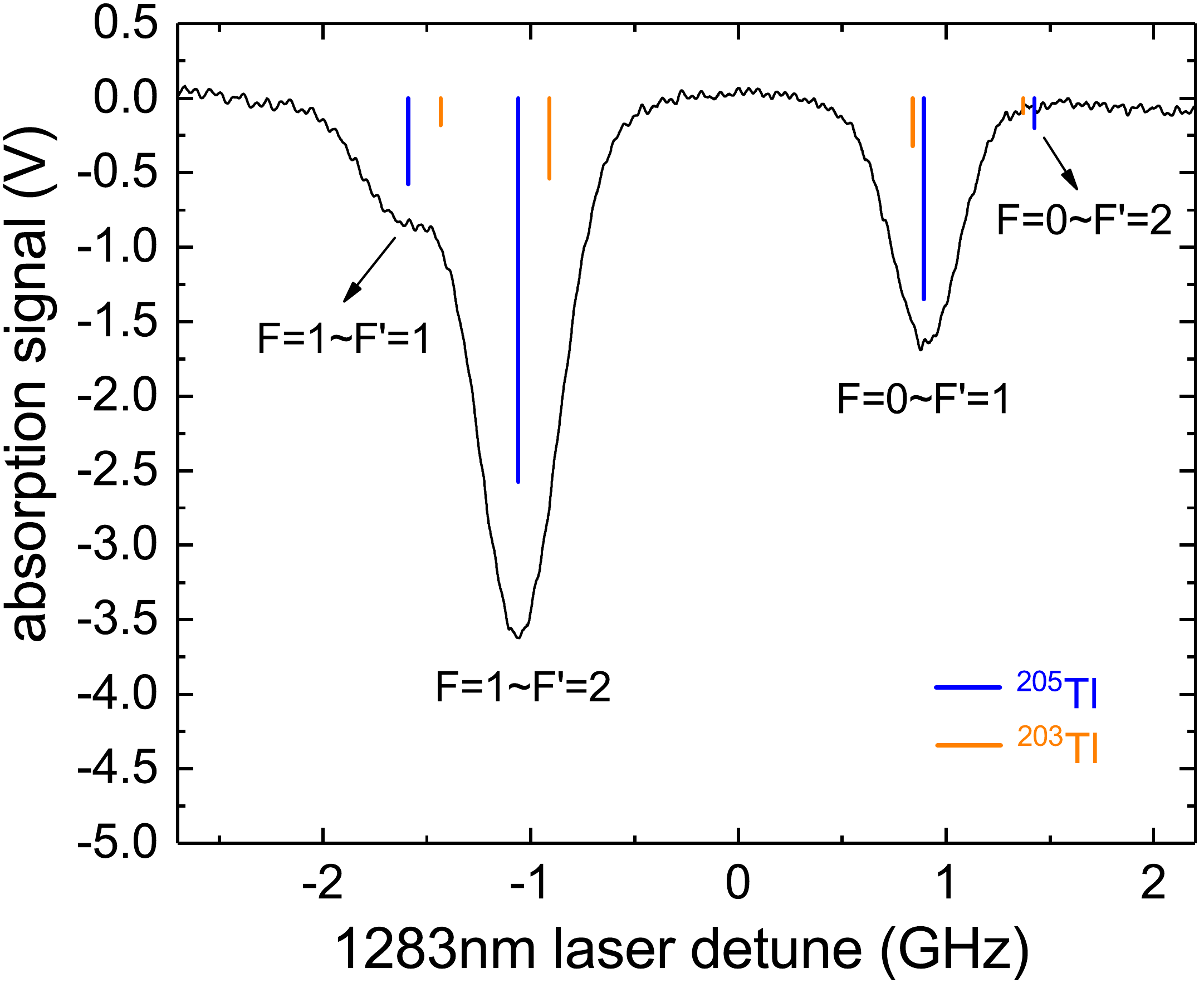}
\caption{The Tl $6P_{1/2}$ to $6P_{3/2}$ absorption signal. 
This spectrum was taken using the SAM technique with 9.36~GHz sidebands to bridge the large frequency gap and to avoid overlapping.
The hyperfine transitions for both $^{205}$Tl and $^{203}$Tl are all indicated.}
\label{1283nmspectrum}
\end{figure}

Figure~\ref{1283nmspectrum} shows the direct absorption spectrum of $6P_{1/2}\rightarrow6P_{3/2}$ forbidden transition using the 1283~nm QD-ECDL laser.
These transitions are all $M1$-allowed except the $F=0\rightarrow F'=2$ transition, which is a much weaker $E2$-allowed transition.
We utilized the sideband amplitude modulation (SAM) technique \cite{chen2019sideband} to improve the signal to noise ratio (SNR), and no intensity variation within the scanning range was found.
The SAM could effectively eliminate the power fluctuation and maintain the absorption profile in the spectrum. 
In the meantime, it also bridged the largely separated hyperfine structure and allowed us to observe the entire spectrum by scanning the 1283~nm with only $\sim5$~GHz for a 21.3~GHz frequency gap.
With a cell temperature at 620~$^\circ$C, the observed Doppler width was 344~MHz. 
Such a broadening made the $6P_{3/2}$ hyperfine splitting incompletely resolvable, and all the small isotopic transitions of $^{203}$Tl were buried under the strong transitions.

\subsubsection{535~nm coupling transition}

\begin{figure}
\centering
\includegraphics[width=300 pt]{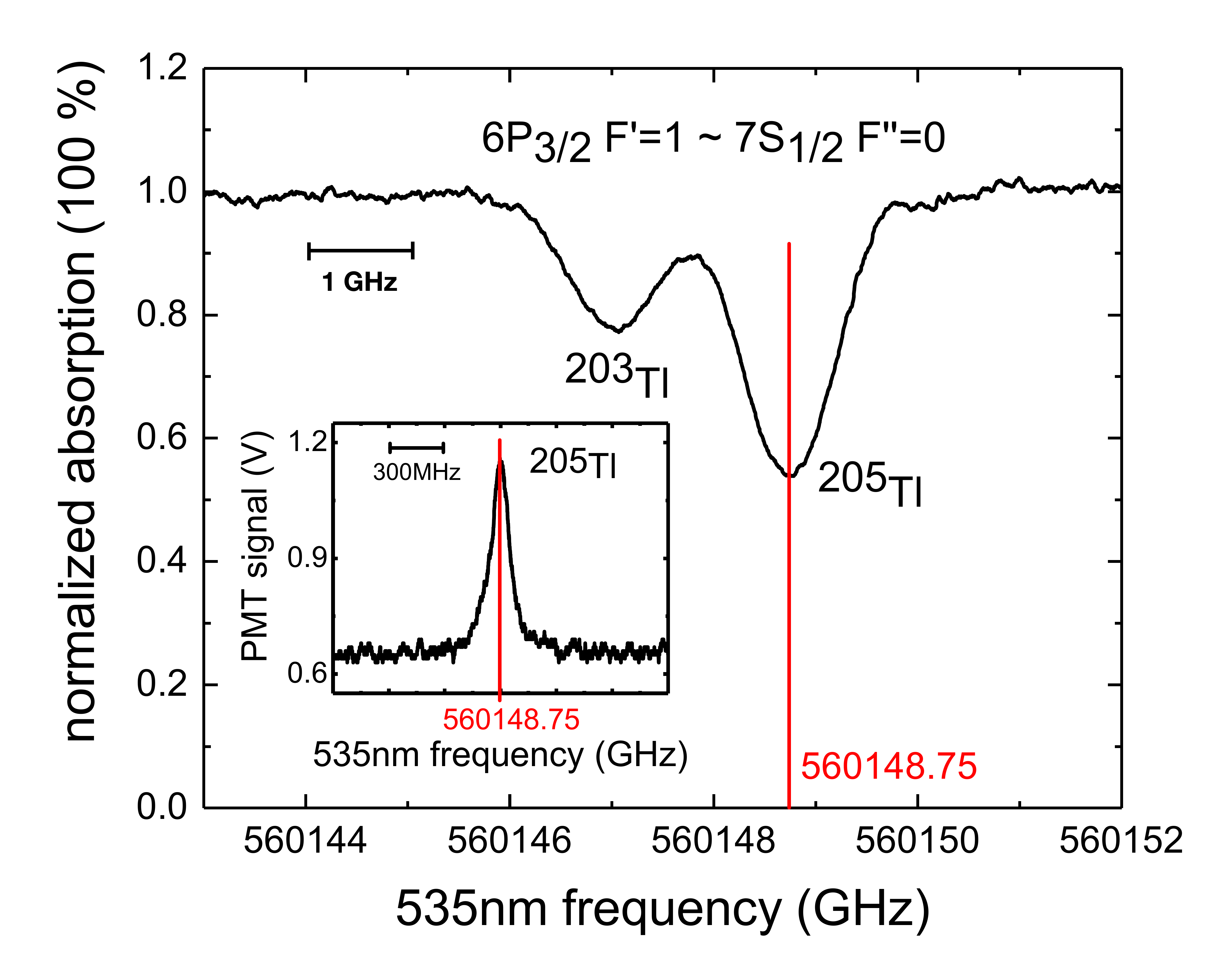}
\caption{The Tl $6P_{3/2}(F=1)$ to $7S_{1/2}(F=0)$ absorption signal using amplitude modulation by an AOM.
The isotope shifts are clearly distinguished. 
The inset is the fluorescence spectrum using the atomic beam system.
The linewidth is 110~MHz.}
\label{535nmspectrum}
\end{figure}

Figure~\ref{535nmspectrum} shows the absorption signals of the $6P_{3/2}(F=1)\rightarrow7S_{1/2}(F=0)$ transition of both $^{205}$Tl and $^{203}$Tl.
These transitions start from the metastable $6P_{3/2}$ state, which is thermally populated by the high cell temperature (620~$^\circ$C).
In this transition, the isotope shift is larger than the Doppler width, so the isotopic transition can be completely resolved.
The Doppler-broadened linewidth is $\sim1$~GHz.
The laser frequency was measured using a wavemeter with 1~MHz resolution. 
This coupling laser was locked to the resonance of $6P_{3/2}(F=1)\rightarrow7S_{1/2}(F=0)$ to have $\Delta_{\rm{c}}=0$ (Eq.~(\ref{rho12})) while performing the EIT spectroscopy. 
With $\Delta_{\rm{c}}\neq0$, the EIT profile is asymmetry and results in a systematic error in $\chi$, as the theoretical model shows.
To avoid the asymmetry, we cross-checked the central frequency with our atomic beam apparatus as a secondary calibration, which has no Doppler broadening as the inset of Fig.~\ref{535nmspectrum} shows.
In our atomic beam spectroscopy, the uncertainty of the absolute frequency was estimated to be 3~MHz from the SNR, and the linewidth was observed to be 110~MHz.
The systematic error of this absolute frequency due to the asymmetrical profile was \textless 3~MHz.
The measured transition frequencies of $6P_{3/2}\rightarrow7S_{1/2}$ using the atomic beam and a wavemeter were in good agreement with the Doppler free saturation spectroscopy \cite{shie2013frequency}.
In the $\chi$ measurement, we utilized computer feedback control to stabilize the coupling laser frequency for $\Delta_{\rm{c}}=0$ by forwarding the wavemeter output.
The long term stability was 1~MHz in 1~min. 
By choosing a different target frequency for locking, we were able to select different isotopes for the $\chi$ measurement and were allowed to measure the $\chi$ of $^{203}$Tl.

\section{EIT spectrum}

\begin{figure}
\centering
\includegraphics[width=290 pt]{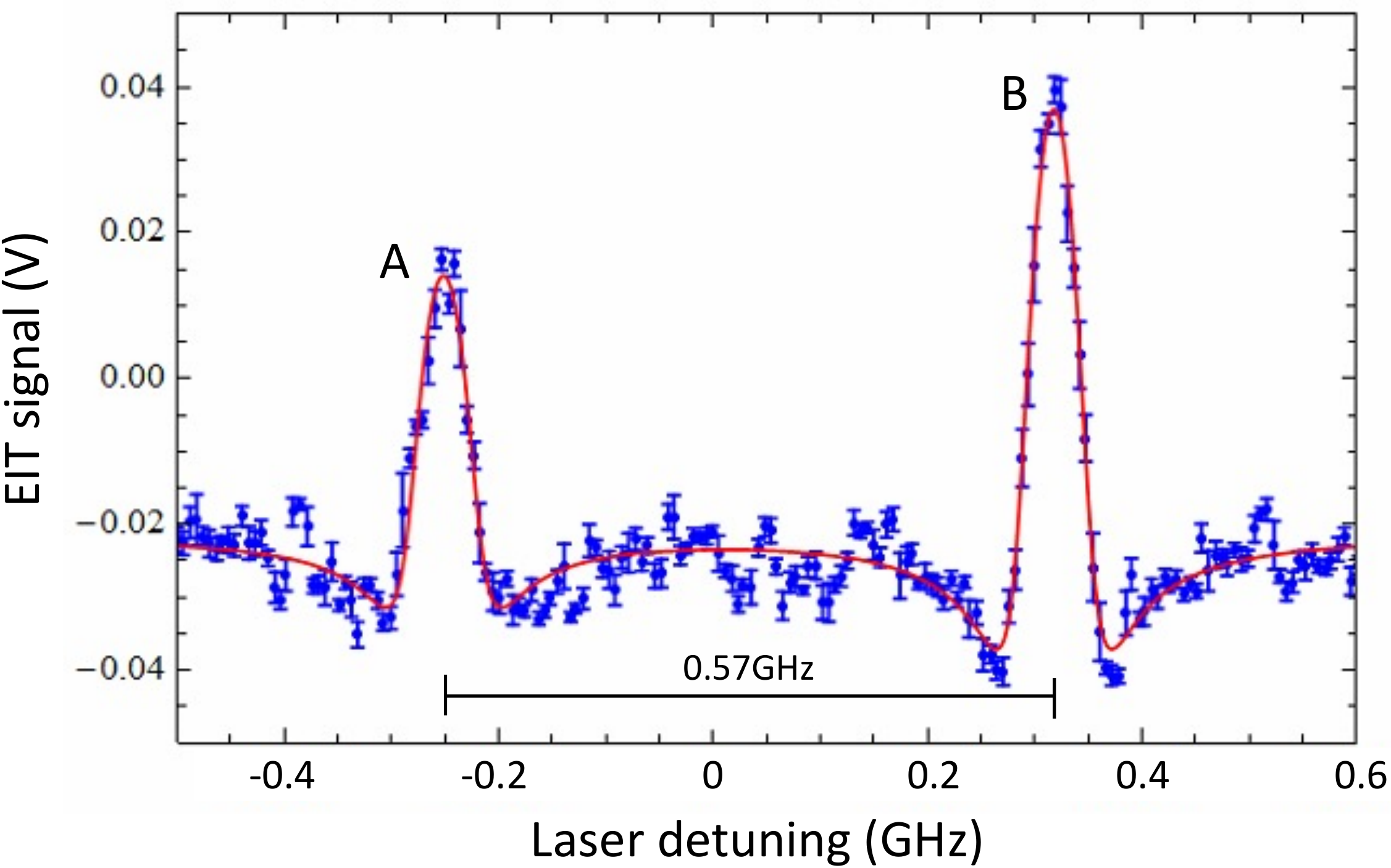}
\caption{The EIT spectrum of channel~A and channel~B. 
The blue points and error bars are the experimental results with data binning for every 50 points. 
The red curve is the fitted ladder-type EIT model. }
\label{fittingresidual}
\end{figure}

To perform the EIT spectroscopy for $\chi$ measurement, we stabilized the coupling laser as we mentioned earlier, and we scanned the 1283~nm probe laser through the two channels with various polarization angles relative to the coupling beam.
The atomic cell temperature was $577~^\circ$C.
Fig.~\ref{fittingresidual} shows a typical experimental run in 20 seconds at one polarization angle, where the signal amplitude of channel~A and channel~B was approximately equal. 
Data binning was made for every 50 data points. 
The frequency axis was calibrated using a reference cavity with FSR=1.4~GHz, and the spacing of two transition signals was 569.5~MHz because of the hyperfine splitting of $6P_{1/2}$ (21.3~GHz) and the applied f-EOM sideband gap (20.74~GHz).
The signal was from the output of the lock-in amplifier with a time constant of 30~ms. 

\begin{figure}
\centering
\includegraphics[width=300 pt]{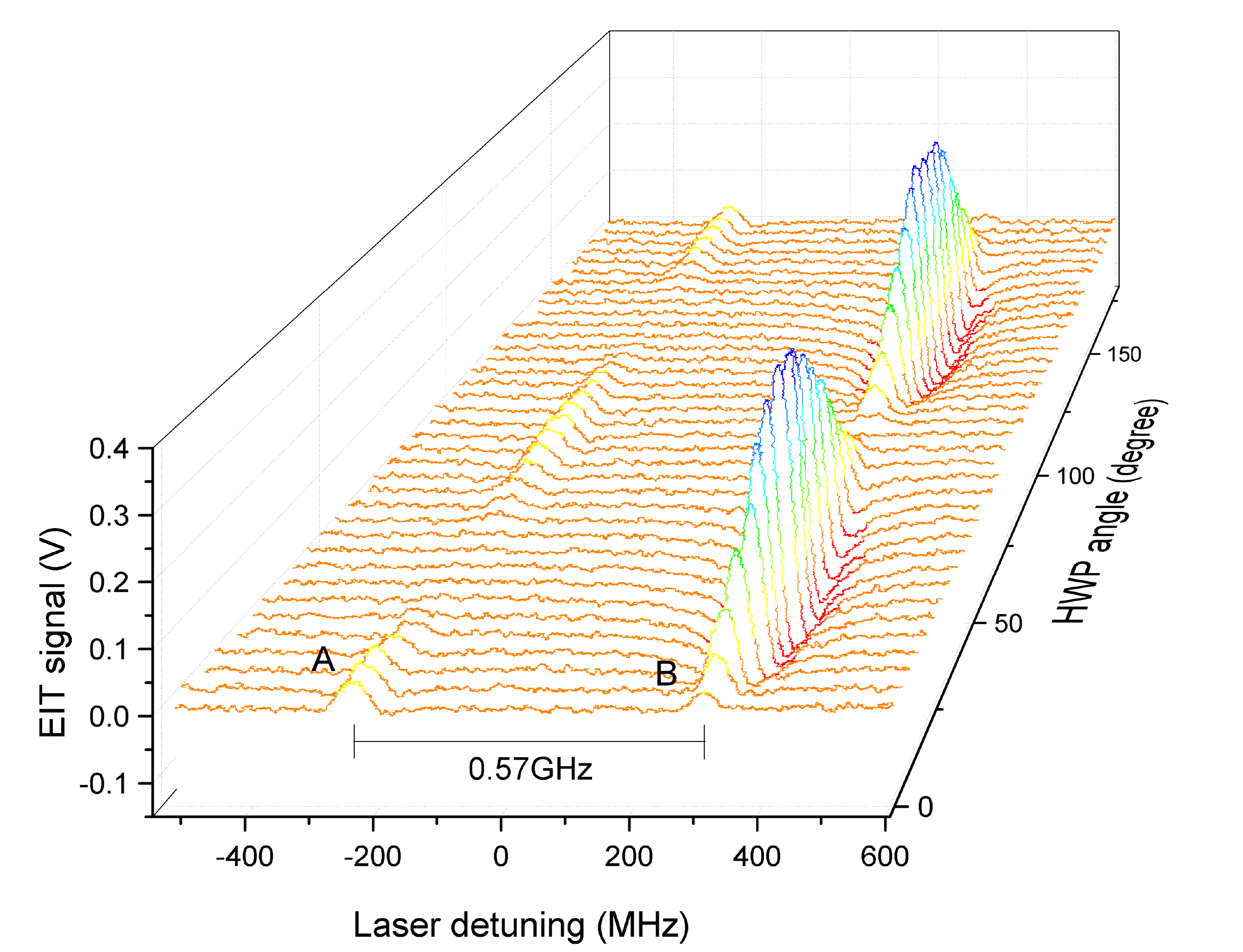}
\caption{The channel~A and channel~B spectrum with polarization angles from 0 to $2\pi$.
The spacing between channel~A and channel~B was only 0.57~GHz because of the f-EOM bridging technique. 
The $\pi/2$ phase shift between the two channels was observed.}
\label{3dcompare}
\end{figure}

The background noise level was $\sim6.1$~mV.
While the signal of one channel was maximum, the other channel was buried under the noise level as Fig.~\ref{3dcompare} shows.
The $\chi$ measurement was mainly inferred from the amplitude of two channels varied with the polarization angle. 
The noise of the spectrum was the primary uncertainty source.
For a single run, the uncertainty of the $\chi$ value could be estimated by the ratio of the maximum amplitude to the noise level.

Figure~\ref{fittingresidual} also exemplifies the model fitting to the experimental data. 
The red curve is the Doppler-broadened model (Eq.~(\ref{eta})), which gives $S_{\rm{A}}$ ($S_{\rm{B}}$) as 2.7 (4.4)~V with 10\% uncertainty, $\gamma_{\rm{A}}$ ($\gamma_{\rm{B}}$) as $7.7(3)\times10^6$ ($4.3(3)\times10^7$)~sec$^{-1}$, and $\Omega_{\rm{c}}$ as $2\pi\times10(2)$~MHz.
The 10\% uncertainty of the signal amplitude was the result of the background noise. 
The fitted value of $\gamma_{\rm{A}}$ ($\gamma_{\rm{B}}$) is not just the lifetime decay rate of the corresponding states because of the laser frequency instability and scanning jitter.
Meanwhile, the strong collision quench should be taken into account.
These broadenings (for $\gamma_{\rm{A, B}}$) are common for both channel~A and channel~B.
Therefore, their impact on the $\chi$ value is minimized in our differential-type measurement.
The coupling strength ($\Omega_{\rm{c}}$) agrees with the calculation from the $6P_{3/2}\rightarrow7S_{1/2}$ dipole transition \cite{safronova2005excitation} and the applied coupling laser intensity.

Figure~\ref{3dcompare} shows the spectrum of two EIT signals with various polarization angles (0 to $2\pi$) between the probe and the coupling beams. 
The spacing of the two transition signals was 569.5~MHz to prevent the two EIT signals from interfering with each other. 
The channel~A signal was the maximum while the two polarizations were in parallel, whereas the channel~B signal was the maximum in the condition of the perpendicular polarization. 

\begin{figure}
\centering
\includegraphics[width=300 pt]{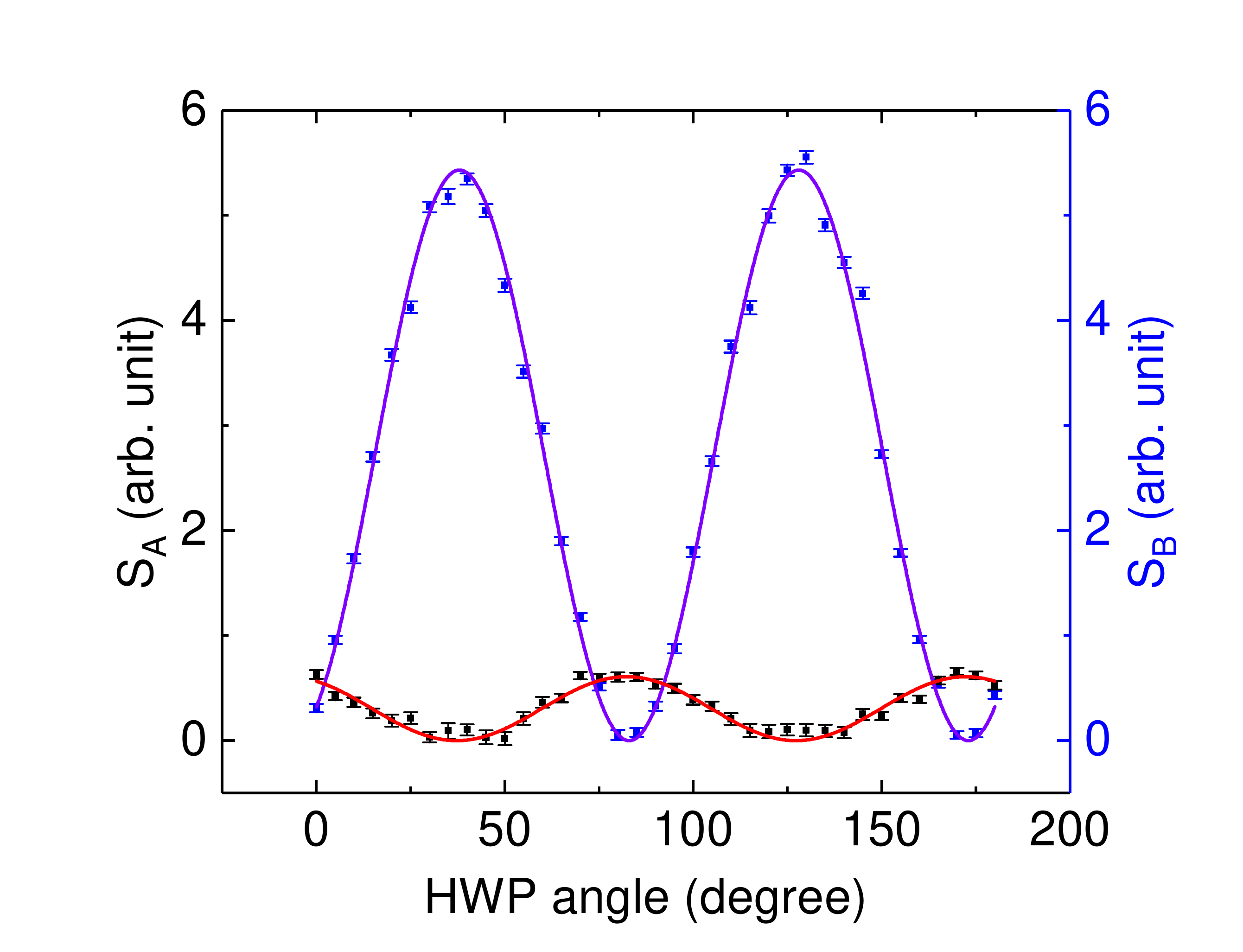}
\caption{The fitted signal amplitudes ($S_{\rm{A}}$ and $S_{\rm{B}}$) versus the polarization angles. 
The black and blue points are the values of $S_{\rm{A}}$ and $S_{\rm{B}}$, respectively.
Some amplitudes are too small to be above the noise level.
The red line and purple line are fitting curves, corresponding to the squared sinusoidal functions with a $\phi/2$ phase difference. 
The error bars are the model parameter uncertainties ($S_{\rm{A,B}}$) given by the fitting in Fig.~\ref{fittingresidual}.
HWP: half-wave plate. The HWP angle varies from 0 to $\pi$, which is equivalent to the polarization angles from 0 to $2\pi$.}
\label{cossinsquare}
\end{figure}

The signal amplitudes are proportional to the probe transition moments and functions of polarization angles, Eq.~(\ref{cossin}). 
Thus, the amplitudes, $S_{\rm{A,B}}(\theta)$, from the fitted EIT profile were plotted in Fig.~\ref{cossinsquare} for one of the experimental runs with the polarization angle varying from 0 to $2\pi$.
The polarization dependent amplitude could be expressed as:
\begin{equation}
\label{phi}
\begin{split}
    S_{\rm{A}}(\theta)=&~S_{\rm{A0}}\cos^2{(\theta+\phi_{\rm{A}})}\\
   S_{\rm{B}}(\theta)=&~S_{\rm{B0}}\sin^2{(\theta+\phi_{\rm{B}})},\\
\end{split}
\end{equation}
where $\theta/2$ is the HWP angle relative to the rotation angle gauge, $\phi_{\rm{A,B}}$ are offset angles for each channel. 
The amplitude of channel~A (B) is $S_{\rm{A0}}$ ($S_{\rm{B0}}$), which presents the maximum amplitude among all polarization angles.
In Fig.~\ref{cossinsquare}, the deviations of each data point mainly came from the noise of the EIT spectrum as Fig.~\ref{fittingresidual} shows.
In this example run, $S_{\rm{A0}}=0.61(1)$ and $S_{\rm{B0}}=5.43(4)$ were used as one data point to calculate the $S_{\rm{A0}}/S_{\rm{B0}}=(1/4)|(3\chi/\sqrt{5})-1|^2$. 
In the average of all experimental runs, $\phi_{\rm{A}}-\phi_{\rm{B}}=0.061(2)^\circ$ which implied that the maximum amplitude of the two signals has a $\pi/2$ phase difference in their polarization angle as the theory predicted.

\section{Systematic effects and error budget}

In this section, we discuss the systematic errors and estimate their contribution to the inferred $\chi$ value as Table~\ref{error} lists. 
These assessments include the coupling laser frequency uncertainty, the cell temperature, the laser power fluctuation, and the probe laser scanning nonlinearity.

\begin{center}
\begin{table}[b]
\caption{\label{error}The error budget of systematic effects in $\chi$.}
\centering
\begin{tabular}{cccc}
\br
\textrm{Systematic error}&
\textrm{relative uncertainty}\\
 & ($10^{-5}$)\\
\mr
535nm frequency lock uncertainty & $0.8$\\
Temperature & $9.0$\\
Laser power variation & $8.8$\\
Unequal sidebands & $255.0$\\
Residual magnetic field & $9.5$\\
Laser frequency scanning speed variation & $94.1$\\
\mr
Total & $272.3$\\
\br
\end{tabular}
\end{table}
\end{center}

\subsection{Light power effect}

The theoretical model (Eq.~(\ref{rho12})) used in our experiment was under the \emph{weak probe approximation}, which implied that the signal amplitude was proportional to the square of the probe laser Rabi frequency ($4{\Omega_{\rm{a,b}}}^2\propto I_p$) and independent of the coupling laser power ($4{\Omega_{\rm{c}}}^2\propto I_{\rm{coupling}}$).
We have examined the applicability by varying the probe and the coupling power.

\begin{figure}[htbp]
\centering
\includegraphics[width=300 pt]{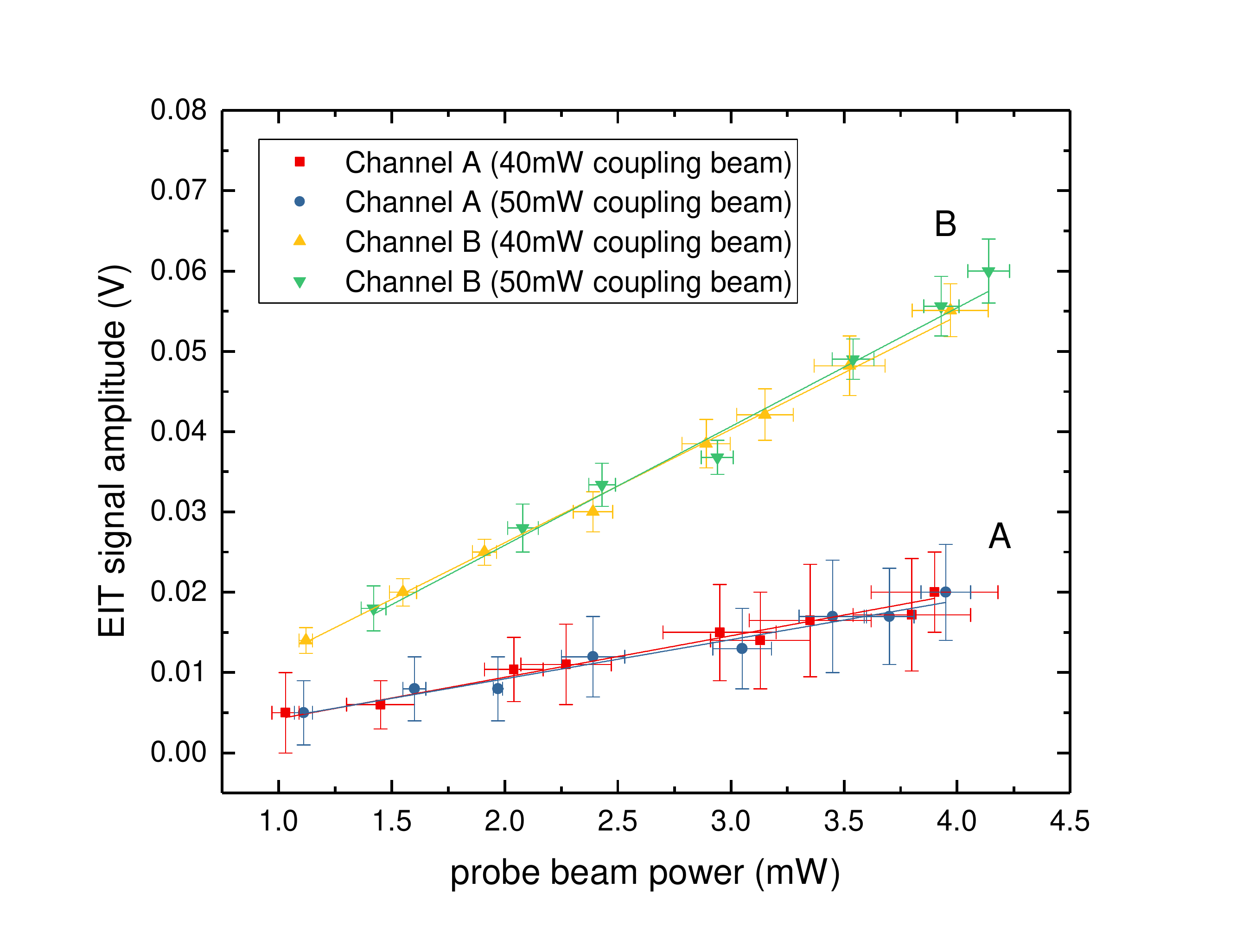}
\caption{The channel A and channel B signals at various probe beam power. 
The data points are the repeating measurements with various coupling and probe power.
And, the error bars of both x and y represent one standard deviation.
During one measurement, the fluctuation of the probe power gave errors in the x-axis.
The slopes of the linear fittings are $5.2(3)\times10^{-3}$ (red), $4.9(4)\times10^{-3}$ (blue), $1.41(3)\times10^{-2}$ (yellow), and $1.48(9)\times10^{-2}$ (green).
In this figure, even with a potential 25\% coupling power variation, our measurement is still well under “weak problem approximation”.
The channel A signals used the same photodetector but with 3x gain.}
\label{power}
\end{figure}

The EIT amplitudes versus the probe powers ($\Omega_{\rm{a,b}}^2$) with different coupling powers (${\Omega_{\rm{c}}}^2$) are shown in Fig.~\ref{power}. 
It verifies that our experimental setting was well under the weak probe approximation, and the model was eligible.
In our experiment, the typical coupling power was 50~mW and the fluctuation was 1.7\%. 
As the result shows, such small fluctuation was negligible in the resulting signal amplitude measurement.

\subsection{Number density}

\begin{figure}
\centering
\includegraphics[width=300 pt]{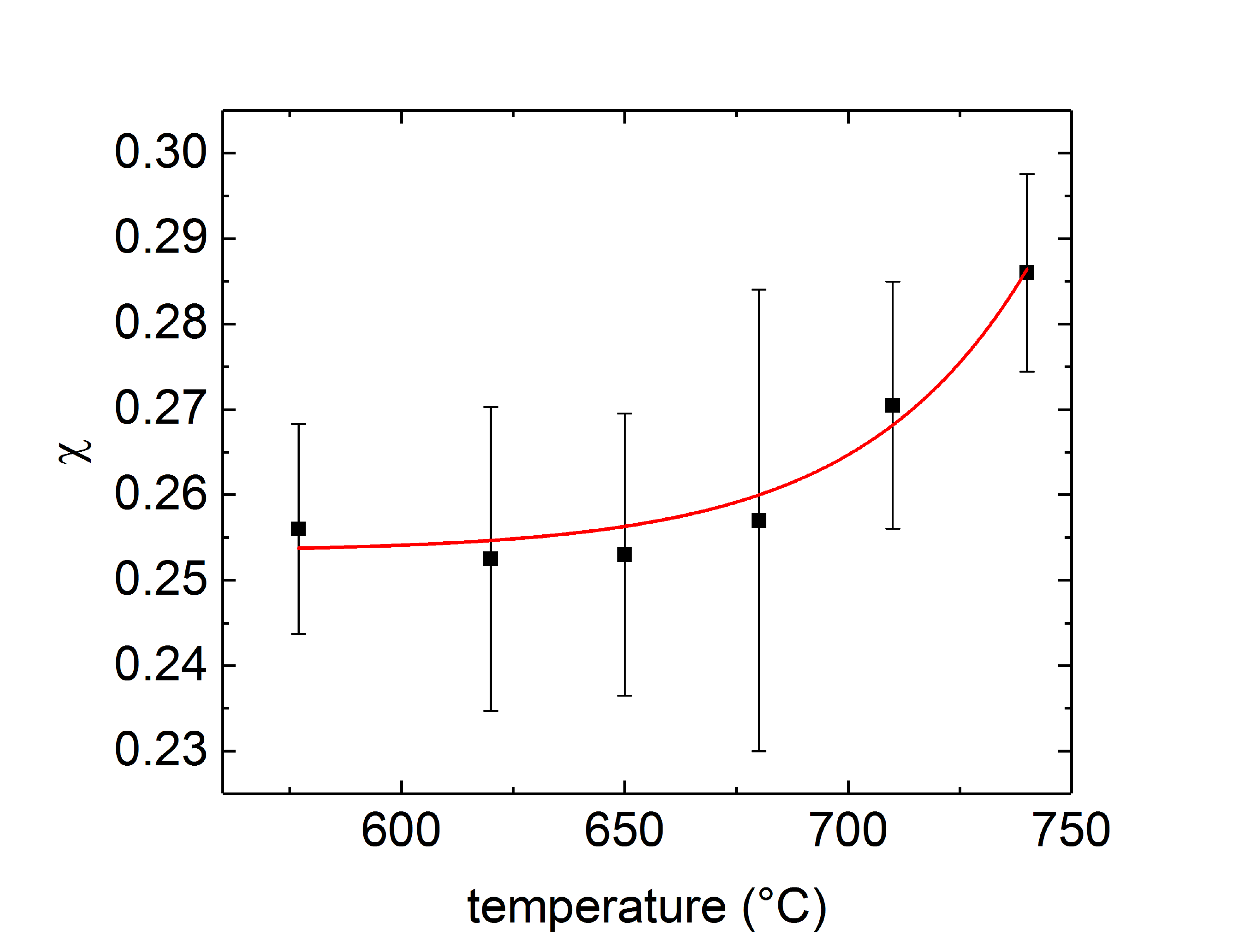}
\caption{The $\chi$ value at different temperatures of the thallium cell. 
Because the coupling beam power became dimmed along its propagation while it passed through the optically thick cell at higher temperatures, it probably caused the signal ratio of channel~A to channel~B to change, and so did the $\chi$ value. 
The red line shows the trend of the $\chi$ value rising at higher temperatures.
The error bars of $\chi$ in 1$\sigma$ uncertainty calculated (propagating) from the uncertainties of $S_{\rm{A0,B0}}$ from Fig.~\ref{cossinsquare}, then calculated using Eq.~(\ref{cossin}) through the standard error propagating procedure.}
\label{chitotemperature}
\end{figure}

The number density not only affected the amplitude of the signal but also resulted in an inhomogeneous coupling laser intensity distribution along the laser beam while the number density was sufficiently high. 
Because the coupling transition was $E1$-allowed with a large interaction cross-section, the absorption lowered the coupling laser power along the laser propagation. 
Thus, the cell was optically thick at a high temperature. 
This effect has not been included in our model and led to a systematic error in the measurement. 
There was a trade off in the high number density despite providing a larger signal. 
We inferred the $\chi$ value under various cell temperatures (various number densities) and found that the $\chi$ value increased significantly when the cell temperature was higher as Fig.~\ref{chitotemperature} shows. 
Thus, to avoid such an error in the final data taking, we set the cell at $577~^\circ$C, where the SNR was still sufficient.

\subsection{Error budget}

One of the systematic errors was the frequency of the stabilized coupling beam.
Firstly, the frequency jitter uncertainty of the 535~nm coupling laser was estimated to be 1~MHz because the feedback control frequency stabilization was based on the high finesse wavemeter, which has a relative uncertainty of 1~MHz.
The lineshape, $L(\Delta_1,\Delta_{\rm{c}};\Omega_{\rm{c}})$, changed by the $\Delta_{\rm{c}}$ frequency jitter, could result in signal amplitude fluctuation.
The signature of such noise was that it is more pronounced on the peak of the signal but not on the signal baseline.
As Fig.~\ref{fittingresidual} shows, the noise on the top of the EIT signal is as large as the background noise. 
We concluded that the frequency jitter was negligible since the linewidth was as large as 75~MHz.
Secondly, the nonzero coupling laser detune ($\Delta_{\rm{c}}\neq0$) resulted in an asymmetrical EIT profile as we mentioned earlier. 
This frequency ($\Delta_{\rm{c}}$) was manually set to be on resonance ($\Delta_{\rm{c}}=0$) within a deviation (\textless~3~MHz). 
We simulated the EIT profile with $\Delta_{\rm{c}}=\pm3$~MHz in comparison with the experimental data.
The relative deviation on the derived $\chi$ was only $0.8\times10^{-5}$. Such a small effect was thanks to the differential measurement, since both channel~A and channel~B shared the same coupling beam and therefore the same induced asymmetrical profile.

The cell temperature fluctuation affected the Maxwell-Boltzmann distribution and changed the density of thallium atoms. 
In our experiments, the cell temperature was stabilized to be $\delta T$~\textless~$0.5~^{\circ}$C/hr, which was corresponding to a 1.3\% fluctuation in the number density and 0.03\% in the Doppler width.
Because the time gap between scanning through channel~A and channel~B was only 13~secs, the relative fluctuation of the EIT profile amplitude was only $9\times10^{-5}$.
The variation of the Doppler width was related to the $\gamma_{\rm{s}}'$ variation and had a small impact on the $\chi$ value.

Both the intensities of the coupling and the probe beams were also the sources of uncertainty and errors.
The power variation effect of the coupling beam was negligible as we discussed earlier.
The power instability of the probe laser could be categorized into two parts: the fast intensity noise and the slow intensity fluctuation. 
The fast noise was mostly filtered out by the 30~ms time constant of the lock-in amplifier and further averaged out by the binning average.
The slow intensity fluctuation was reduced using the frequency bridging and the intensity normalization.

By applying the f-EOM bridging technique, the observation time interval between channel~A and channel~B signals was only 13~secs. 
The deviation of the laser power was 1.1\% during this interval. Thus, it was important to normalize the EIT signal with respect to the probe power, which was monitored by a detector (PD3 in Fig.~\ref{expsetup}).
The residual normalization error due to the noise of the power monitor signal (0.8\%) was estimated to be $1.1\%\times0.8\%=8.8\times10^{-5}$.

While the applied modulation to the EOM is a perfect sinusoidal wave, the power equality between two sidebands is guaranteed by the momentum conservation.
However, certain systematic effects can violate this symmetry. 
Firstly, the Serrodyne modulation \cite{wong1982electro,johnson1988serrodyne,johnson2010broadband} generates the unequal sidebands by a sawtooth wave, which is an asymmetrical microwave with high order harmonics.
The ratio of the rising- to falling-times determines the ratio of the two sideband powers.
To achieve a pair of equal power sidebands, the high harmonics, particularly the second harmonic $2\times10.37$~GHz, were suppressed. The 10.37~GHz microwave modulation was generated from 1.296~GHz from a synthesizer (HP-8648B), then multiplied ($\times8$) using three multipliers (ZX90-2-19-S+, ZX90-2-36-S+, ZX90-2-24-S+) and filtered by one bandpass filter (VBFZ-5500-S+).
Overall, the lowest (second) harmonic was suppressed by $-20$~dB.
Then, it was amplified by a 10 GHz amplifier, which further cleans out the high frequency components. 
Finally, the f-EOM (PM-0K5-10-PFU-PFU-130) has a 10~GHz cut-off frequency.
Therefore, we estimated the suppression of the second harmonic $2\times10.37$~GHz \textless~$-30$~dB, the corresponding rising- and falling-times asymmetry \textless~$0.13\%$, and the inequality between two sidebands \textless~$255.0\times10^{-5}$ \cite{johnson1988serrodyne,li2010optical}.

Secondly, “time-varying” unequal sideband, that is, residual amplitude modulation (RAM) is considered \cite{whittaker1985residual,bi2019suppressing}.
While the input polarization was not aligned to the axis of EOM (or PM fiber of the f-EOM), any polarization-sensitive component could result in such a RAM, which is a fast amplitude modulation with a frequency of 10.37 GHz.
In the experiments, our InGaAs detector could only detect signals from DC to 100 kHz, and our lock-in amplifier time constant was 30 ms. 
The RAM (10.37~GHz) effect was completely averaged out.
Even though, we had carefully adjusted the polarization to minimize RAM.
We conclude that RAM has no effect on our measurement.

As discussed in \cite{cronin1998studies,marangos1998electromagnetically,cheng2017electromagnetically}, the stray magnetic field would distort the EIT signals, $L(\Delta_1,\Delta_{\rm{c}};\Omega_{\rm{c}})$, because of the Zeeman splitting. 
In our experiment, the residual magnetic field was measured to be 0.2~G, and the corresponding Zeeman splittings were 93~kHz for the $6P_{1/2}(F=1)$ state and 0.56~MHz for the $6P_{3/2}(F'=1)$ state. 
The splittings of both states are much narrower than the width of the EIT signal which is $\sim75$~MHz.  
Experimentally, we increased the magnetic field up to 2~G, and no broadening or change of the lineshape was observed.
A small systematic uncertainty due to the dichroism was then estimated.
The differential method in calculating the $\chi$ value also reduced the impact of the Zeeman splitting of the $6P_{3/2}(F'=1)$ state (the common state).
However, for the $6P_{1/2}$ state, the Zeeman splitting could not be eliminated, because channel~A from $6P_{1/2}(F=1)$ suffered the Zeeman splitting, but not channel~B from $6P_{1/2}(F=0)$. 
To assess such an effect from the residual magnetic field, we modeled the Zeeman splitting using two theoretical profiles to compose one EIT signal. 
The resulting deviation of $\chi$ was estimated to be $9.5\times10^{-5}$ in relative uncertainty.

The probe laser scanning speed fluctuation caused variation of the EIT profile widths, but did not directly affect the amplitude, due to the slow scanning speed (1.1~GHz in 25~s) and 30~ms time constant of the lock-in amplifier. 
In our 32 experimental runs, the resulting $\gamma_{\rm{A, B}}$ for each run had a 12\% deviation according to our frequency marker (the reference cavity with 1.4~GHz FSR). 
As $\chi$ was derived from the amplitude ratio of channel~A to channel~B, the indirect effect of the $\gamma_{\rm{A, B}}$ variation on the amplitude ratio, $\chi$, was estimated to be $94.1\times10^{-5}$ using our theoretical model.

Benefiting from the differential method in calculating the $\chi$ value, most systematic errors could be eliminated because of the same impacts on channel A and channel B. 
The most severe systematic uncertainties were the laser frequency scanning speed variation and the unequal sidebands (\textless~0.3\%). 
All the systematic errors are listed in Table~\ref{error}. 
The total systematic uncertainty was $272.3\times10^{-5}$.

\section{Result and Conclusion}

\subsection{$\chi$ value for $^{205}$Tl and $^{203}$Tl}

\begin{figure}
\centering
\includegraphics[width=300 pt]{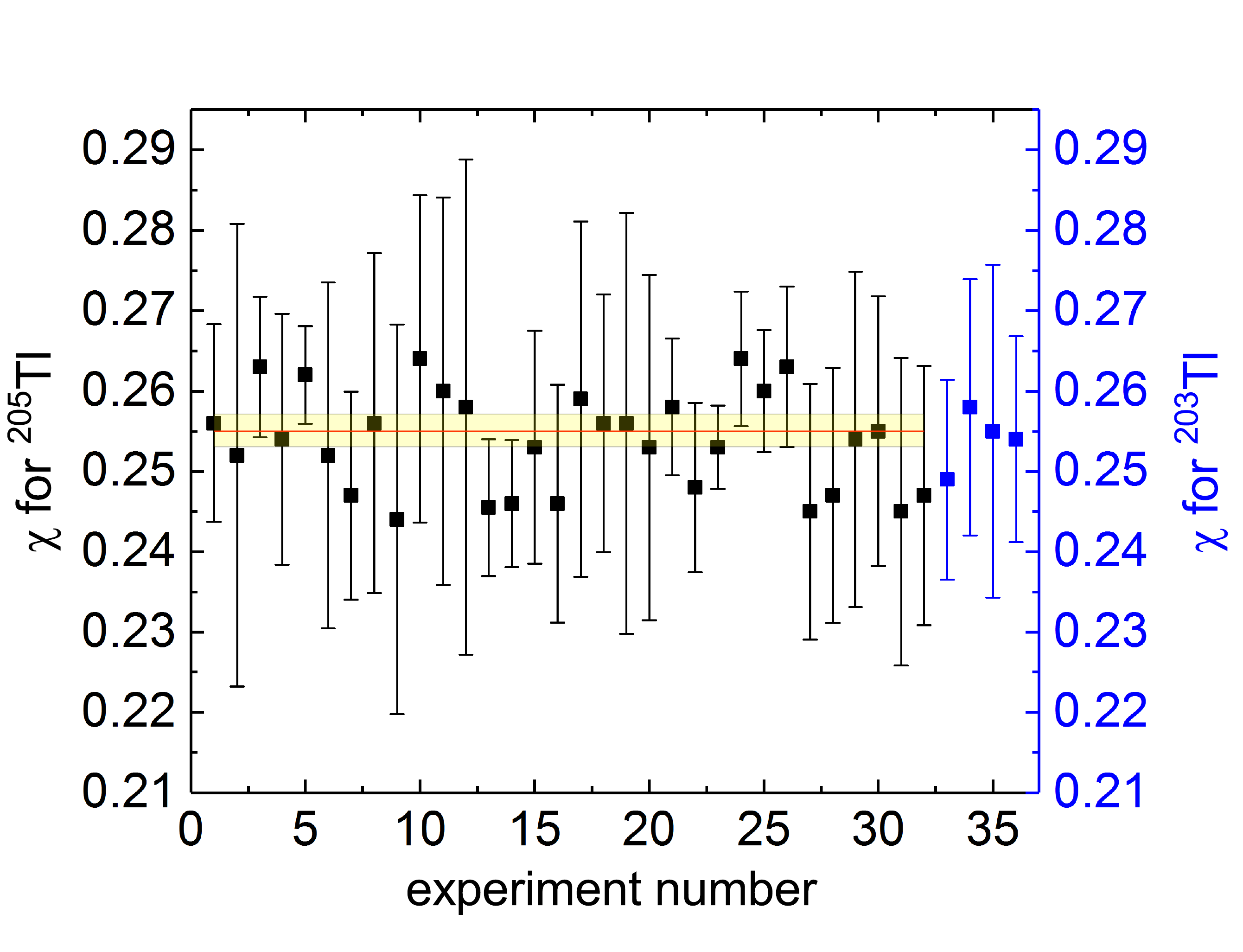}
\caption{The experimental results with $^{205}$Tl and $^{203}$Tl. 
The black points are 32 experimental runs with $^{205}$Tl. 
The red line is the average $\chi$ value for $^{205}$Tl. 
The blue points are 4 experimental runs with $^{203}$Tl.
The error bars are also calculated (propagating) from the uncertainties of $S_{\rm{A0,B0}}$.}
\label{expresult}
\end{figure}

For the final $\chi$ value measurement, we have performed 32 data taking runs. 
Each run was a set of EIT spectrum with the polarization angle scanning through $2\pi$. 
These data were taken over a period of 6 days. Each spectrum was fitted to the model to extract the relevant parameters. 
Then, the $\chi$ value could be inferred from each run, which shows as one data point in Fig.~\ref{expresult}. 
Each run gave a $\chi$ value with 8\% relative uncertainty. The final statistical error was 0.8\%, shown as central yellow $1\sigma$ area in Fig.~\ref{expresult}. 
The $\chi$ value of $^{205}$Tl was given as:
\begin{equation}
    \chi_{205}=0.2550(20)(7)
\end{equation}
By tuning the coupling laser frequency, we were able to pick out $^{203}$Tl for the isotopic $\chi$ value measurement. 
The result was shown in the last 4 data points on the right side of Fig.~\ref{expresult}, which gives:
\begin{equation}
    \chi_{203}=0.2532(73)(7)
\end{equation}

Our result is in good agreement with one of the theoretical calculations, $\chi=0.254$ \cite{neuffer1977calculation}.
The comparison of all the calculations and experiments is listed in Table~\ref{previous chi}. 
The source of the discrepancy is unclear.

\begin{table}[t]
\caption{\label{previous chi}Previous experimental and theoretical values of $\chi$ for Tl.}
\centering
\begin{tabular}{m{12em} m{2.5cm} m{1cm}}
\br
Theory & \\
\mr
1964\cite{garstang1964transition} & 0.21\\
1968\cite{warner1968transition} & 0.219\\
1977\cite{neuffer1977calculation} & 0.254\\
1995\cite{maartensson1995magnetic} & 0.2401\\
1996\cite{biemont1996forbidden} & 0.278\\
2005\cite{safronova2005excitation} & 0.2369\\
\br
Experiment & \\
\mr
1995\cite{vetter1995precise} & 0.240(2)\\
1998\cite{cronin1998studies} & 0.22\\
1999\cite{majumder1999measurement} & 0.2387(10)(38)\\
This experiment ($^{205}$Tl) & 0.2550(20)(7)\\
This experiment ($^{203}$Tl) & 0.2532(73)(7)\\
\br
\end{tabular}
\end{table}

\subsection{Connection between $\chi$ and thallium PNC}

As the analysis in \cite{majumder1999measurement}, a part of the $2.2\sigma$ discrepancy of the $R_{PNC}$ measurements between the Seattle group and the Oxford group can be attributed to the 6\% overestimate of the $\chi$ results.
The $\chi$ value, which indirectly determines the $R_{PNC}$ in the Oxford experiment, affects the lineshape parameters such as component linewidths and optical depth.
However, the Seattle group used independent Faraday rotation calibration, so their measurement is less sensitive to any newly measured $\chi$ value.
With our measurement ($\chi=0.2550$), which is close to the value ($\chi=0.254$) that was used in the Oxford experiment, the discrepancy of $R_{PNC}$ between the two groups is still not resolved as \cite{majumder1999measurement}.
The source of this important discrepancy remains unclear.

Our measured $\chi$ value, the amplitude of electrical quadruple amplitude, from the sub-Doppler EIT profile with the frequency bridging technique disagrees with the previous experimental results. 
The $\chi$ value is an important ingredient in the optical rotation PNC experiments and a stringent test of the wave function calculation for atomic thallium.
With our $\chi$ measurement, the PNC measurement for isotopes, which the atomic cesium lacks, provides an opportunity to measure the NSD effect.
The EIT spectroscopy can also be applied to the PNC measurement as demonstrated in \cite{cronin1998studies}. 
Using a laser source with high intensity stability could achieve a low noise signal and avoid the trade-off of the high number density.

\ack{This work was financially supported by the Ministry of Science and Technology (MOST, Taiwan) and the Center for Quantum Technology from the Featured Areas Research Center Program within the framework of the Higher Education Sprout Project by the Ministry of Education (MOE, Taiwan), Grand No. MOST-109-2112-M-007-020-MY3 and MOST-110-2634-F-007-022.}

\newcommand{\newblock}{}


\begin{thebibliography}{10}

\bibitem{bouchiat1974weak}
MA~Bouchiat and CC~Bouchiat.
\newblock Weak neutral currents in atomic physics.
\newblock {\em Physics Letters B}, 48(2):111--114, 1974.

\bibitem{bouchiat2011atomic}
Marie-Anne Bouchiat.
\newblock Atomic parity violation. early days, present results, prospects.
\newblock {\em arXiv preprint arXiv:1111.2172}, 2011.

\bibitem{sandars1987parity}
PGH Sandars.
\newblock Parity and time-reversal violation in atoms and molecules.
\newblock {\em Physica Scripta}, 36(6):904, 1987.

\bibitem{dzuba2012revisiting}
VA~Dzuba, JC~Berengut, VV~Flambaum, and B~Roberts.
\newblock Revisiting parity nonconservation in cesium.
\newblock {\em Physical review letters}, 109(20):203003, 2012.

\bibitem{porsev2009precision}
SG~Porsev, K~Beloy, and A~Derevianko.
\newblock Precision determination of electroweak coupling from atomic parity
  violation and implications for particle physics.
\newblock {\em Physical review letters}, 102(18):181601, 2009.

\bibitem{geetha1998nuclear}
KP~Geetha, Angom~Dilip Singh, BP~Das, and CS~Unnikrishnan.
\newblock Nuclear-spin-dependent parity-nonconserving transitions in ba+ and
  ra+.
\newblock {\em Physical Review A}, 58(1):R16, 1998.

\bibitem{ginges2004violations}
JSM Ginges and Victor~V Flambaum.
\newblock Violations of fundamental symmetries in atoms and tests of
  unification theories of elementary particles.
\newblock {\em Physics Reports}, 397(2):63--154, 2004.

\bibitem{sahoo2021new}
BK~Sahoo, BP~Das, and H~Spiesberger.
\newblock New physics constraints from atomic parity violation in cs 133.
\newblock {\em Physical Review D}, 103(11):L111303, 2021.

\bibitem{flambaum1997anapole}
VV~Flambaum and DW~Murray.
\newblock Anapole moment and nucleon weak interactions.
\newblock {\em Physical Review C}, 56(3):1641, 1997.

\bibitem{flambaum1984nuclear}
VV~Flambaum, IB~Khriplovich, and OP~Sushkov.
\newblock Nuclear anapole moments.
\newblock {\em Physics Letters B}, 146(6):367--369, 1984.

\bibitem{dzuba2011calculation}
VA~Dzuba and VV~Flambaum.
\newblock Calculation of nuclear-spin-dependent parity nonconservation in s--d
  transitions of ba+, yb+, and ra+ ions.
\newblock {\em Physical Review A}, 83(5):052513, 2011.

\bibitem{dmitriev2004p}
VF~Dmitriev and IB~Khriplovich.
\newblock P and t odd nuclear moments.
\newblock {\em Physics reports}, 391(3-6):243--260, 2004.

\bibitem{haxton2002nuclear}
WC~Haxton, C-P Liu, and Michael~J Ramsey-Musolf.
\newblock Nuclear anapole moments.
\newblock {\em Physical Review C}, 65(4):045502, 2002.

\bibitem{wood1997measurement}
CS~Wood, SC~Bennett, Donghyun Cho, BP~Masterson, JL~Roberts, CE~Tanner, and
  Carl~E Wieman.
\newblock Measurement of parity nonconservation and an anapole moment in
  cesium.
\newblock {\em Science}, 275(5307):1759--1763, 1997.

\bibitem{dzuba2002high}
VA~Dzuba, VV~Flambaum, and JSM Ginges.
\newblock High-precision calculation of parity nonconservation in cesium and
  test of the standard model.
\newblock {\em Physical Review D}, 66(7):076013, 2002.

\bibitem{sahoo2011parity}
BK~Sahoo, P~Mandal, and M~Mukherjee.
\newblock Parity nonconservation in odd isotopes of single trapped atomic ions.
\newblock {\em Physical Review A}, 83(3):030502, 2011.

\bibitem{sahoo2011parity2}
BK~Sahoo and BP~Das.
\newblock Parity nonconservation in ytterbium ion.
\newblock {\em Physical Review A}, 84(1):010502, 2011.

\bibitem{edwards1995precise}
NH~Edwards, SJ~Phipp, PEG Baird, and S~Nakayama.
\newblock Precise measurement of parity nonconserving optical rotation in
  atomic thallium.
\newblock {\em Physical review letters}, 74(14):2654, 1995.

\bibitem{vetter1995precise}
PA~Vetter, DM~Meekhof, PK~Majumder, SK~Lamoreaux, and EN~Fortson.
\newblock Precise test of electroweak theory from a new measurement of parity
  nonconservation in atomic thallium.
\newblock {\em Physical review letters}, 74(14):2658, 1995.

\bibitem{dzuba1987calculation}
VA~Dzuba, VV~Flambaum, PG~Silvestrov, and OP~Sushkov.
\newblock Calculation of parity non-conservation in thallium.
\newblock {\em Journal of Physics B: Atomic and Molecular Physics (1968-1987)},
  20(14):3297, 1987.

\bibitem{haxton2001atomic}
WC~Haxton and Carl~E Wieman.
\newblock Atomic parity nonconservation and nuclear anapole moments.
\newblock {\em Annual Review of Nuclear and Particle Science}, 51(1):261--293,
  2001.

\bibitem{majumder1999measurement}
PK~Majumder and Leo~L Tsai.
\newblock Measurement of the electric quadrupole amplitude within the 1283-nm 6
  p 1/2- 6 p 3/2 transition in atomic thallium.
\newblock {\em Physical Review A}, 60(1):267, 1999.

\bibitem{safronova2005excitation}
UI~Safronova, MS~Safronova, and WR~Johnson.
\newblock Excitation energies, hyperfine constants, e 1, e 2, and m 1
  transition rates, and lifetimes of 6 s 2 n l states in tl i and pb ii.
\newblock {\em Physical Review A}, 71(5):052506, 2005.

\bibitem{porsev2012electric}
SG~Porsev, MS~Safronova, and MG~Kozlov.
\newblock Electric dipole moment enhancement factor of thallium.
\newblock {\em Physical Review Letters}, 108(17):173001, 2012.

\bibitem{tang2018relativistic}
Yong-Bo Tang, Ning-Ning Gao, Bing-Qiong Lou, and Ting-Yun Shi.
\newblock Relativistic coupled-cluster calculations of the polarizabilities of
  atomic thallium.
\newblock {\em Physical Review A}, 98(6):062511, 2018.

\bibitem{maartensson1995magnetic}
Ann-Marie M{\aa}rtensson-Pendrill.
\newblock Magnetic moment distributions in tl nuclei.
\newblock {\em Physical review letters}, 74(12):2184, 1995.

\bibitem{garstang1964transition}
RH~Garstang.
\newblock Transition probabilities of forbidden lines.
\newblock {\em Journal of Research of the National Bureau of Standards. Section
  A, Physics and Chemistry}, 68(1):61, 1964.

\bibitem{warner1968transition}
BRIAN Warner.
\newblock Transition probabilities in np and n p5configurations.
\newblock {\em Zeitschrift fur Astrophysik}, 69:399, 1968.

\bibitem{neuffer1977calculation}
David~V Neuffer and Eugene~D Commins.
\newblock Calculation of parity-nonconserving effects in the 6 p 1 2 2- 7 p 1 2
  2 forbidden m 1 transition in thallium.
\newblock {\em Physical Review A}, 16(3):844, 1977.

\bibitem{biemont1996forbidden}
Emile Bi{\'e}mont and Pascal Quinet.
\newblock Forbidden lines in 6pk (k= 1--5) configurations.
\newblock {\em Physica Scripta}, 54(1):36, 1996.

\bibitem{wolfenden1991observation}
TD~Wolfenden, PEG Baird, and PGH Sandars.
\newblock Observation of parity-violating optical rotation in atomic thallium.
\newblock {\em EPL (Europhysics Letters)}, 15(7):731, 1991.

\bibitem{macpherson1991precise}
MJD Macpherson, KP~Zetie, RB~Warrington, DN~Stacey, and JP~Hoare.
\newblock Precise measurement of parity nonconserving optical rotation at 876
  nm in atomic bismuth.
\newblock {\em Physical review letters}, 67(20):2784, 1991.

\bibitem{meekhof1993high}
Dawn~M Meekhof, P~Vetter, PK~Majumder, SK~Lamoreaux, and EN~Fortson.
\newblock High-precision measurement of parity nonconserving optical rotation
  in atomic lead.
\newblock {\em Physical review letters}, 71(21):3442, 1993.

\bibitem{cronin1999new}
Alexander~Douglas Cronin.
\newblock {\em New techniques for measuring atomic parity violation}.
\newblock University of Washington, 1999.

\bibitem{cronin1998studies}
AD~Cronin, RB~Warrington, SK~Lamoreaux, and EN~Fortson.
\newblock Studies of electromagnetically induced transparency in thallium vapor
  and possible utility for measuring atomic parity nonconservation.
\newblock {\em Physical review letters}, 80(17):3719, 1998.

\bibitem{dzuba1986enhancement}
VA~Dzuba, VV~Flambaum, and IB~Khriplovich.
\newblock Enhancement ofp-andt-nonconserving effects in rare-earth atoms.
\newblock {\em Zeitschrift f{\"u}r Physik D Atoms, Molecules and Clusters},
  1(3):243--245, 1986.

\bibitem{pollock1992atomic}
SJ~Pollock, E~Norval Fortson, and L~Wilets.
\newblock Atomic parity nonconservation: Electroweak parameters and nuclear
  structure.
\newblock {\em Physical Review C}, 46(6):2587, 1992.

\bibitem{safronova2018search}
MS~Safronova, D~Budker, D~DeMille, Derek F~Jackson Kimball, A~Derevianko, and
  Charles~W Clark.
\newblock Search for new physics with atoms and molecules.
\newblock {\em Reviews of Modern Physics}, 90(2):025008, 2018.

\bibitem{antypas2019isotopic}
Dionysis Antypas, A~Fabricant, Jason~Evan Stalnaker, Konstantin Tsigutkin,
  VV~Flambaum, and Dmitry Budker.
\newblock Isotopic variation of parity violation in atomic ytterbium.
\newblock {\em Nature Physics}, 15(2):120--123, 2019.

\bibitem{fortson1990nuclear}
EN~Fortson, Y~Pang, and L~Wilets.
\newblock Nuclear-structure effects in atomic parity nonconservation.
\newblock {\em Physical review letters}, 65(23):2857, 1990.

\bibitem{brown2009calculations}
B~Alex Brown, A~Derevianko, and VV~Flambaum.
\newblock Calculations of the neutron skin and its effect in atomic parity
  violation.
\newblock {\em Physical Review C}, 79(3):035501, 2009.

\bibitem{shie2013frequency}
Nang-Chian Shie, Chun-Yu Chang, Wen-Feng Hsieh, Yi-Wei Liu, and Jow-Tsong Shy.
\newblock Frequency measurement of the 6 p 3/2→ 7 s 1/2 transition of
  thallium.
\newblock {\em Physical Review A}, 88(6):062513, 2013.

\bibitem{chen2012absolute}
Tzu-Ling Chen, Isaac Fan, Hsuan-Chen Chen, Chang-Yi Lin, Shih-En Chen,
  Jow-Tsong Shy, and Yi-Wei Liu.
\newblock Absolute frequency measurement of the 6 p 1/2→ 7 s 1/2 transition
  in thallium.
\newblock {\em Physical Review A}, 86(5):052524, 2012.

\bibitem{joshi2003electromagnetically}
Amitabh Joshi and Min Xiao.
\newblock Electromagnetically induced transparency and its dispersion
  properties in a four-level inverted-y atomic system.
\newblock {\em Physics Letters A}, 317(5-6):370--377, 2003.

\bibitem{khan2016role}
Sumanta Khan, Vineet Bharti, and Vasant Natarajan.
\newblock Role of dressed-state interference in electromagnetically induced
  transparency.
\newblock {\em Physics Letters A}, 380(48):4100--4104, 2016.

\bibitem{zhu1996sub}
Yifu Zhu and TN~Wasserlauf.
\newblock Sub-doppler linewidth with electromagnetically induced transparency
  in rubidium atoms.
\newblock {\em Physical Review A}, 54(4):3653, 1996.

\bibitem{gea1995electromagnetically}
Julio Gea-Banacloche, Yong-qing Li, Shao-zheng Jin, and Min Xiao.
\newblock Electromagnetically induced transparency in ladder-type
  inhomogeneously broadened media: Theory and experiment.
\newblock {\em Physical Review A}, 51(1):576, 1995.

\bibitem{grimm2000optical}
Rudolf Grimm, Matthias Weidem{\"u}ller, and Yurii~B Ovchinnikov.
\newblock Optical dipole traps for neutral atoms.
\newblock In {\em Advances in atomic, molecular, and optical physics},
  volume~42, pages 95--170. Elsevier, 2000.

\bibitem{chen2017sub}
Tzu-Ling Chen and Yi-Wei Liu.
\newblock Sub-doppler resolution near-infrared spectroscopy at 1.28 $\mu$m with
  the noise-immune cavity-enhanced optical heterodyne molecular spectroscopy
  method.
\newblock {\em Optics letters}, 42(13):2447--2450, 2017.

\bibitem{chen2019sideband}
Wei-Ling Chen, Tzu-Ling Chen, and Yi-Wei Liu.
\newblock Sideband amplitude modulation absorption spectroscopy of ch 4 at 1170
  nm.
\newblock {\em Optics Express}, 27(15):21264--21272, 2019.

\bibitem{wong1982electro}
KK~Wong, RM~De~La~Rue, and S~Wright.
\newblock Electro-optic-waveguide frequency translator in linbo 3 fabricated by
  proton exchange.
\newblock {\em Optics letters}, 7(11):546--548, 1982.

\bibitem{johnson1988serrodyne}
Leonard~M Johnson and Ch~H Cox.
\newblock Serrodyne optical frequency translation with high sideband
  suppression.
\newblock {\em Journal of lightwave technology}, 6(1):109--112, 1988.

\bibitem{johnson2010broadband}
DMS Johnson, JM~Hogan, S-W Chiow, and MA~Kasevich.
\newblock Broadband optical serrodyne frequency shifting.
\newblock {\em Optics letters}, 35(5):745--747, 2010.

\bibitem{li2010optical}
Yanlu Li, Stijn Meersman, and Roel Baets.
\newblock Optical frequency shifter on soi using thermo-optic serrodyne
  modulation.
\newblock In {\em 7th IEEE International Conference on Group IV Photonics},
  pages 75--77. IEEE, 2010.
  
  \bibitem{whittaker1985residual}
Edward~A Whittaker, Manfred Gehrtz, and Gary~C Bjorklund.
\newblock Residual amplitude modulation in laser electro-optic phase
  modulation.
\newblock {\em JOSA B}, 2(8):1320--1326, 1985.

\bibitem{bi2019suppressing}
Jin Bi, Yunlin Zhi, Liufeng Li, and Lisheng Chen.
\newblock Suppressing residual amplitude modulation to the 10- 7 level in
  optical phase modulation.
\newblock {\em Applied Optics}, 58(3):690--694, 2019.

\bibitem{marangos1998electromagnetically}
Jonathan~P Marangos.
\newblock Electromagnetically induced transparency.
\newblock {\em Journal of modern optics}, 45(3):471--503, 1998.

\bibitem{cheng2017electromagnetically}
Hong Cheng, Han-Mu Wang, Shan-Shan Zhang, Pei-Pei Xin, Jun Luo, and Hong-Ping
  Liu.
\newblock Electromagnetically induced transparency of 87rb in a buffer gas cell
  with magnetic field.
\newblock {\em Journal of Physics B: Atomic, Molecular and Optical Physics},
  50(9):095401, 2017.

\end{thebibliography}
\end{document}